\documentclass[11pt]{article}

\usepackage{epsfig}
\usepackage{theorem}
\usepackage{enumerate}
\usepackage{amsmath}
\usepackage{amssymb}
\usepackage{boxedminipage}
\usepackage{subfigure}
\usepackage{url}
\usepackage{color}
\usepackage{epsfig}

\newcommand{\titlestring}{Finding Ground States of Sherrington-Kirkpatrick Spin Glasses with Hierarchical BOA and Genetic Algorithms}
\newcommand{\reportnumber}{2008004}
\newcommand{\shortauthors}{Martin Pelikan, Helmut G. Katzgraber and Sigismund Kobe}
\newcommand{\datestring}{January 2008}
\definecolor{myblue}{rgb}{0.165,0.34,0.5}
\date{}

\begin{document}

\begin{titlepage}
\setlength{\parindent}{0pt}

\noindent
\includegraphics[width=5in]{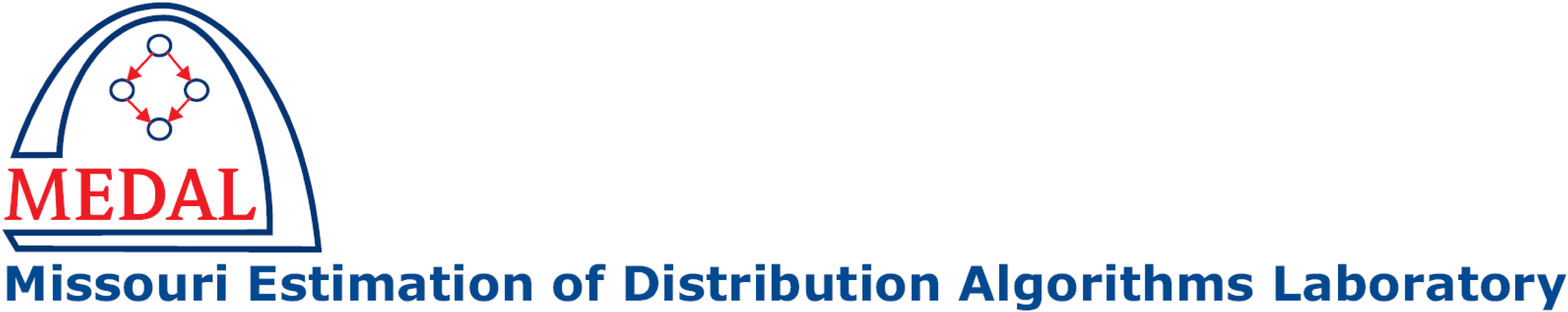}
\vspace*{0.075in}
{\color{myblue}
\hrule height 2pt
}
\vspace*{0.5in}

{\bf
\textsf{{\large
\titlestring}}
}
\vspace*{0.25in}

\textsf{\shortauthors}

\vspace*{0.25in}

\textsf{MEDAL Report No. \reportnumber}

\vspace*{0.25in}

\textsf{\datestring}

\vspace*{0.25in}

{\bf \textsf{Abstract}}  

\vspace*{0.075in}

{\small \textsf{This study focuses on the problem of finding ground states of random
instances of the Sherrington-Kirkpatrick (SK) spin-glass model
with Gaussian couplings. While the ground states of SK spin-glass
instances can be obtained with branch and bound, the computational
complexity of branch and bound yields instances of not more than about
90 spins. We describe several approaches based on the hierarchical
Bayesian optimization algorithm (hBOA) to reliably identifying ground
states of SK instances intractable with branch and bound, and present
a broad range of empirical results on such problem instances. We argue
that the proposed methodology holds a big promise for reliably
solving large SK spin-glass instances to optimality with practical time
complexity. The proposed approaches to identifying global optima reliably
can also be applied to other problems and they can be used with many
other evolutionary algorithms. Performance of hBOA is compared to
that of the genetic algorithm with two common crossover operators.
}}


\vspace*{0.25in}

{\bf \textsf{Keywords}}

\vspace*{0.075in}

{\small \textsf{Sherrington-Kirkpatrick spin glass, hierarchical BOA,
genetic algorithm, estimation of distribution algorithms, evolutionary
computation, branch and bound.}}

\vfill

\noindent
\begin{minipage}{6in}
{\small \textsf{Missouri Estimation of Distribution Algorithms 
Laboratory (MEDAL)\\
Department of Mathematics and Computer Science\\
University of Missouri--St. Louis\\
One University Blvd.,
St. Louis, MO 63121\\
E-mail: \url{medal@cs.umsl.edu}\\
WWW: \url{http://medal.cs.umsl.edu/}\\}}
\end{minipage}

\end{titlepage}

\title{\titlestring}

\author{
{\bf Martin Pelikan}\\
Missouri Estimation of Distribution Algorithms Laboratory (MEDAL)\\
Dept. of Math and Computer Science, 320 CCB\\
University of Missouri at St. Louis\\
One University Blvd., St. Louis, MO 63121\\
\url{pelikan@cs.umsl.edu}
\and
{\bf Helmut G. Katzgraber}\\
Theoretische Physik, ETH Z\"{u}rich\\
Schafmattstrasse 32\\
CH-8093 Z\"{u}rich, Switzerland\\
\url{katzgraber@phys.ethz.ch}
\and
{\bf Sigismund Kobe}\\
Technische Universit\"{a}t Dresden\\
Institut f\"{u}r Theoretische Physik\\
01062 Dresden, Germany\\
\url{kobe@physik.tu-dresden.de}
}

\maketitle


\begin{abstract}
This study focuses on the problem of finding ground states of random
instances of the Sherrington-Kirkpatrick (SK) spin-glass model
with Gaussian couplings. While the ground states of SK spin-glass
instances can be obtained with branch and bound, the computational
complexity of branch and bound yields instances of not more than about
90 spins. We describe several approaches based on the hierarchical
Bayesian optimization algorithm (hBOA) to reliably identifying ground
states of SK instances intractable with branch and bound, and present
a broad range of empirical results on such problem instances. We argue
that the proposed methodology holds a big promise for reliably
solving large SK spin-glass instances to optimality with practical time
complexity. The proposed approaches to identifying global optima reliably
can also be applied to other problems and they can be used with many
other evolutionary algorithms. Performance of hBOA is compared to
that of the genetic algorithm with two common crossover operators.
\end{abstract}

\noindent



\section{Introduction}

Spin glasses are prototypical models for disordered systems, which
provide a rich source of challenging theoretical and computational
problems. Despite ongoing research over the last two to three decades
little is known about the nature of the spin-glass state, in particular
at low temperatures \cite{mezard:87,binder:86,kawashima:03}. Hence
numerical studies at low and zero temperature are of paramount
importance.  One of the most interesting but also most numerically
challenging spin-glass models is the Sherrington-Kirkpatrick
(SK) spin glass~\cite{Sherrington:78}, in which interactions
between spins have infinite range. The model has the advantage
that analytical solutions exist \cite{parisi:79,talagrand:06}
for certain low-temperature properties. Still, certain properties
of the model are poorly understood.  For example, the behavior
of the ground-state energy fluctuations as a function of the
number of spins is still a source of controversy: Crisanti {\em et
al.} \cite{crisanti:92} found $\rho = 5/6$, whereas Bouchaud {\em et al.}
\cite{bouchaud:03} and Aspelmeier {\em et al.} \cite{aspelmeier:02}
found $\rho = 3/4$.  Numerically, for small system sizes, Katzgraber
{\em et al.} \cite{katzgraber:04c,koerner:06} have found $\rho =
0.775(2)$ in agreement with results of Palassini.\cite{palassini:03a}
The aforementioned results by Katzgraber {\em et al.}~have recently
been criticized by Billoire \cite{billoire:07} due to the use of too
small system sizes. Therefore, it would be of interest to estimate
the ground-state energy fluctuations for large system sizes using
reliable ground-state instances.

In optimization, even spin glasses arranged on finite-dimensional
lattices have proven to represent an interesting
class of challenging problems and various evolutionary
algorithms have been shown to locate ground states of large
finite-dimensional spin-glass instances efficiently and reliably
\cite{Hartmann:96,Hartmann:01,Pelikan:06}. Nonetheless, only little
work has been done in solving instances of the general SK model,
which is inherently more complex than spin-glass models arranged
on finite-dimensional lattices, such as the two-dimensional (2D)
or three-dimensional (3D) lattice.

This paper applies the hierarchical Bayesian optimization (hBOA)
and the genetic algorithm (GA) to the problem of finding ground
states of random instances of the SK spin glass with Ising spins
and Gaussian couplings. Nonetheless, the proposed approach can be
readily applied to other variants of the SK spin-glass model. Several
approaches based on hBOA to solving large SK spin-glass instances are
proposed and empirically analyzed. While the paper only presents the
results on systems of up to $n=300$ spins, we argue that the proposed
approach can be scaled to much larger systems, especially with the
use of additional efficiency enhancement techniques. Performance of
hBOA is compared to that of GA with several common variation operators.

The paper is organized as follows. Section~\ref{section-algorithms}
outlines the evolutionary algorithms discussed in this
paper. Section~\ref{section-spin-glass} describes the problem of
finding ground states of SK spin glasses and the branch-and-bound
algorithm, which can be used for finding ground states of small
SK instances. Section~\ref{section-initial-experiments} presents
initial experimental results obtained with hBOA on spin glasses
of up to $n=80$ spins. Section~\ref{section-how-to-bigger}
describes several approaches to reliably identifying ground
states of SK spin-glass instances intractable with branch and
bound. Section~\ref{section-experiments-larger} presents and
discusses experimental results obtained with hBOA on spin glasses
of $n\in[100,300]$. Section~\ref{section-experiments-comparison}
compares performance of hBOA and GA with common variation operators on
SK instances of $n\leq 200$ spins. Section~\ref{section-future-work}
outlines future work. Finally, section~\ref{section-conclusions}
summarizes and concludes the paper.


\section{Algorithms}
\label{section-algorithms}

This section outlines the optimization algorithms discussed in
this paper: (1) the hierarchical Bayesian optimization algorithm
(hBOA)~\cite{Pelikan:01*,Pelikan:book} and (2) the genetic algorithm
(GA)~\cite{Holland:75a,Goldberg:89d}. Additionally, the section
describes the deterministic hill climber (DHC)~\cite{Pelikan:03*},
which is incorporated into both algorithms to improve their
performance. Candidate solutions are assumed to be represented by
binary strings of $n$ bits, although all presented methods can be
easily applied to problems with candidate solutions represented by
fixed-length strings over any finite alphabet.

\subsection{Hierarchical BOA (hBOA)}

The hierarchical Bayesian optimization algorithm
(hBOA)~\cite{Pelikan:01*,Pelikan:book} evolves a population of
candidate solutions. The population is initially generated at random
according to the uniform distribution over all $n$-bit strings. Each
iteration starts by selecting a population of promising solutions
using any common selection method of genetic and evolutionary
algorithms; in this paper, we use binary tournament selection. 
Binary tournament selection selects one solution at a time
by first choosing two
random candidate solutions from the current population and
then selecting the best solution out of this subset. Random tournaments
are repeated until the selected population has the same size as the original
population and thus each candidate solution is expected to participate
in two tournaments. 

New solutions are generated by building a Bayesian network with local
structures~\cite{Chickering:97,Friedman:99} for the selected solutions
and sampling from the probability distribution encoded by the built
Bayesian network. To ensure useful-diversity maintenance, the new
candidate solutions are incorporated into the original population
using restricted tournament replacement (RTR)~\cite{Harik:95a}. The
run is terminated when some user-defined termination criteria are met;
for example, the run may be terminated when a previously specified maximum
number of iterations has been executed.

hBOA is an estimation of distribution algorithm
(EDA)~\cite{Baluja:94,Muhlenbein:96**,Larranaga:02,Pelikan:02}.
EDAs---also called probabilistic model-building genetic algorithms
(PMBGAs)~\cite{Pelikan:02} and iterated density estimation algorithms
(IDEAs)~\cite{Bosman:00*}---differ from genetic algorithms by replacing
standard variation operators of genetic algorithms such as crossover
and mutation by building a probabilistic model of promising solutions
and sampling the built model to generate new candidate solutions.

\subsection{Genetic algorithm (GA)}

The genetic algorithm (GA)~\cite{Holland:75a,Goldberg:89d} also
evolves a population of candidate solutions typically represented
by fixed-length binary strings. The first population is generated at
random.  Each iteration starts by selecting promising solutions from
the current population; also in GA we use binary tournament selection. 
New solutions are created by applying variation
operators to the population of selected solutions. Specifically,
crossover is used to exchange bits and pieces between pairs of
candidate solutions and mutation is used to perturb the resulting
solutions. Here we use two-point crossover or uniform crossover, and
bit-flip mutation~\cite{Goldberg:89d}. The new candidate solutions
are incorporated into the original population using restricted
tournament replacement (RTR)~\cite{Harik:95a}. The run is terminated
when termination criteria are met.

The only difference between the hBOA and GA variants discussed in
this paper is the way in which the selected solutions are processed
to generate new candidate solutions.

\subsection{Deterministic Hill Climber (DHC)}

Incorporating local search often improves efficiency of evolutionary
algorithms. For example, in the problem of finding ground states of
instances of the 2D spin glass, incorporating a simple deterministic
hill climber (DHC) into hBOA leads to a speedup of approximately
a factor 10 with respect to the number of evaluations until
optimum~\cite{Pelikan:book}. That is why we decided to incorporate
DHC into both hBOA and GA also in this study.

DHC takes a candidate solution represented by an $n$-bit binary string
on input. Then, it performs one-bit changes on the solution that lead
to the maximum improvement of solution quality. DHC is terminated
when no single-bit flip improves solution quality and the solution
is thus locally optimal with respect to single-bit flips. Here,
DHC is used to improve every solution in the population before the
evaluation is performed.


\section{Sherrington-Kirkpatrick Spin Glass}
\label{section-spin-glass}

This section describes the Sherrington-Kirkpatrick (SK) spin glass
and the branch-and-bound algorithm, which can be used to find ground
states of SK spin-glass instances of relatively small size.

\subsection{SK Spin Glass}

The Sherrington-Kirkpatrick spin glass~\cite{Sherrington:78}
is described by a set of spins $\{s_i\}$ and a set of couplings
$\{J_{i,j}\}$ between all pairs of spins. Thus, unlike in
finite-dimensional spin-glass models, the SK model does not limit the
range of spin-spin interactions to only neighbors in a lattice. For the
classical Ising model, each spin $s_i$ can be in one of two states:
$s_i=+1$ or $s_i=-1$. Note that this simplification corresponds
to highly anisotropic physical magnetic systems; nevertheless, the
two-state Ising model comprises all basic effects also found in more
realistic models of magnetic systems with more degrees of freedom.

For a set of coupling constants $\{J_{i,j}\}$, and a configuration
of spins $C=\{s_i\}$, the energy can be computed as
\begin{equation}
H(C) = - \sum_{i<j} J_{i,j} s_i s_j .
\end{equation}
The usual task in statistical physics is to integrate a known
function over all possible configurations of spins for given coupling
constants, assuming the Boltzmann distribution of spin configurations.
From the physics point of view, it is also interesting to know the
ground states (spin configurations associated with the minimum
possible energy). Finding extremal energies then corresponds to
sampling the Boltzmann distribution with temperature approaching $T
\rightarrow 0$. The problem of finding ground states is NP-complete
even when the interactions are limited only to neighbors in a 3D
lattice~\cite{Barahona:82}; the SK spin glass is thus certainly
NP-complete (unless we severely restrict couplings, making the
problem simpler).

In order to obtain thermodynamically relevant quantities, all
measurements of a spin-glass system have to be averaged over many
disorder instances of random spin-spin couplings.  Here we
consider random instances of the SK model with couplings generated from
the Gaussian distribution with zero mean and unit variance, $N(0,1)$.

\subsection{Branch and Bound}

The branch-and-bound algorithm for finding ground states of SK
spin-glass instances is based on a total enumeration of the space
of all spin configurations. The space of spin configurations is
explored by parsing a tree in which each level corresponds to one
spin and the subtrees below the nodes at this level correspond to
the different values this spin can obtain (for example, the left
subtree sets the spin to $-1$ and the right subtree sets the spin
to $+1$). To make the enumeration more efficient, branch and bound
uses bounds on the energy to cut large parts of the tree, which can
be proved to not lead to better solutions than the best solution
found. We have tried two versions of the branch-and-bound algorithm
for finding ground states of SK spin glasses. Here we outline the
basic principle of the branch and bound that performed best, which
was adopted from references~\cite{Kobe:78,Kobe:84,Kobe:03}.

To understand the energy bounds and the basic procedure of the
branch-and-bound strategy employed in this paper, let us define a
reduced problem which considers only the first $(n-1)$ spins and all
couplings between these spins. Let us denote the minimum energy for
this reduced system by $f_{n-1}^*$. Let us now consider the problem
of setting the last spin, $s_n$, which was excluded in the reduced
system. We know that the optimal energy $f_n^*$ of the full problem
has to satisfy the following inequality:
\begin{equation}
\label{eq-bound-fn}
f_n^* \geq f_{n-1}^* - \sum_{i=1}^{n-1} |J_{i,n}|,
\end{equation}
because $f_{n-1}^*$ is the minimum energy for the reduced system
of only the first $(n-1)$ spins and the largest decrease of energy
by adding $s_n$ into the reduced system is given by the sum of the
absolute values of all the couplings between $s_n$ and other spins.

Analogously, we can define a reduced system with only the first $j$
spins for any $j\leq n$, and denote the minimum energy for such a
system by $f_j^*$. Then, the bound from equation~\ref{eq-bound-fn}
can be generalized as
\begin{equation}
f_j^* \geq f_{j-1}^* - \sum_{i=1}^{j-1} |J_{i,j}|.
\end{equation}

The branch-and-bound algorithm for an SK spin glass of $n$ spins
$\{s_1,s_2,\ldots, s_n\}$ proceeds by iteratively finding the best
configurations for the first $j=2$ to $n$ spins. For each value of
$j$, the result for $(j-1)$ is used to provide the bounds. Whenever
the current branch can be shown to provide at most as good solutions
as the best solution so far, the branch is cut (and not explored).

While it is somewhat surprising that solving $(n-1)$ problems is
faster than a direct solution of the entire problem, the iterative
branch-and-bound strategy is clearly superior to other alternatives
and allows feasible solutions for much larger SK instances. With
the described approach, ground states of SK instances of up to
approximately 90 spins can be found in practical time on a reasonable
sequential computer.

To make the branch-and-bound algorithm slightly faster, we first
execute several runs of a stochastic hill climbing to provide a
relatively good spin configurations for each reduced problem. The
best result of these runs allows the branch-and-bound technique to
cut more branches.


\section{Initial Experiments}
\label{section-initial-experiments}

This section describes initial experiments. As the first step, we
generated a large number of random problem instances of the SK spin-glass 
model of sizes up to $n=80$ spins and applied the branch-and-bound
algorithm to find the ground states of these instances. Next, hBOA
was applied to all these problem instances and the performance and
parameters of hBOA were analyzed.

\subsection{Preparing the Initial Set of Instances of $n\leq 80$ Spins}

First, we have generated $10^4$ SK ground-state instances for each
problem size from $n=20$ spins to $n=80$ spins with step 2. Branch
and bound has been applied to each of these instances to determine
the true ground state, providing us with a set of 310,000 unique
problem instances of different sizes with known ground states.

The motivation for using so many instances for each problem size and
for increasing the problem size with the step of only 2 was that the
problem difficulty varies significantly across the different instances
and we wanted to obtain accurate statistical information about the
performance of hBOA and its parameters; as a result, it was desirable
to obtain as many different problem instances as possible. Since both
branch and bound as well as hBOA can solve such small SK instances
very fast, generating and testing 10,000 instances for each problem size was
feasible with the computational resources available.


\subsection{hBOA Parameters and Experimental Setup}

To represent spin configurations of $n$ spins, hBOA uses an $n$-bit
binary string where the $i$-th bit determines the state of the
spin $s_i$; $-1$ is represented with a~$0$, $+1$ is represented
with a~$1$. Each candidate solution is assigned fitness according
to the energy of the spin configuration it encodes; specifically,
the fitness is equal to the negative energy of the configuration. In
this manner, maximizing fitness corresponds to minimizing energy and the
maximal fitness corresponds to the ground-state energy of the system.

Some hBOA parameters do not depend much on the problem instance
being solved, and that is why they are typically set to some
default values, which were shown to perform well across a broad
range of problems. To select promising solutions, we use binary
tournament selection. New solutions are incorporated into the
original population using restricted tournament replacement with
window size $w=\max\{n, N/5\}$ where $n$ is the number of bits
in a candidate solution and $N$ is the population size. Bayesian
networks are selected based on the Bayesian-Dirichlet metric
with likelihood equivalence~\cite{Cooper:92,Chickering:97},
which is extended with a penalty term to punish overly complex
models~\cite{Friedman:99,Pelikan:01a*,Pelikan:book}. The complexity
of Bayesian networks used in hBOA was not restricted directly and it
only depends on the scoring metric and the training data.

The best values of two hBOA parameters critically depend on the
problem being solved: (1)~the population size and (2)~the maximum
number of iterations. The maximum number of iterations is typically
set to be proportional to the number of bits in the problem, which
is supported by the domino convergence model for exponentially
scaled problems~\cite{Thierens:98}. Since from some preliminary
experiments it was clear that the number of iterations would be very
small for all problems of $n\leq 80$, typically less than $10$ even
for $n=80$, we set the number of iterations to the number of bits in
the problem. Experiments confirmed that the used bound on the number
of iterations was certainly sufficient.

To set the population size, we have used the bisection
method~\cite{Sastry:01c,Pelikan:book}, which automatically determines
the necessary population size for reliable convergence to the optimum
in $10$ out of $10$ independent runs. This is done for each problem
instance so that the resulting population sizes are as small as
possible, which typically minimizes the execution time. Each run in the
bisection is terminated either when the global optimum (ground state)
has been reached (success), or when the maximum number of iterations
has been exceeded without finding the global optimum (failure).

\subsection{Analysis of hBOA on Instances of $n\leq 80$ Spins}

For each problem instance, after determining an adequate population
size with bisection and making 10 independent runs of hBOA with that
population size, we record four important statistics for these 10
successful runs: (1) the population size, (2) the number of iterations,
(3) the number of evaluations, and (4) the number of single-bit flips
of the local searcher. For each problem size, we thus consider 100,000
successful hBOA runs, yielding a total of 3,100,000 successful hBOA
runs for problems of sizes from $n=20$ to $n=80$. In order to solve
larger problems, especially the results for the population size and
the number of iterations are useful. On the other hand, to analyze
the time complexity of hBOA, the number of evaluations and the number
of flips are most important.

The first step in analyzing the results of hBOA is to identify the
probability distribution that the different observed statistics
follow. By identifying a specific distribution type, the results
of the analysis should be much more accurate and practically
useful. Based on our prior work in spin glasses and preliminary
experiments, there are two distributions that might be applicable:
(1) the log-normal distribution and (2) the generalized extreme value
distribution. For all aforementioned statistics we first estimate the
parameters of both the distributions and then compare these estimates
to the underlying data. The most stable results are obtained with
the log-normal distribution and that is why we have decided to use
log-normal distributions in this and further analyses.

\begin{figure}
\epsfig{file=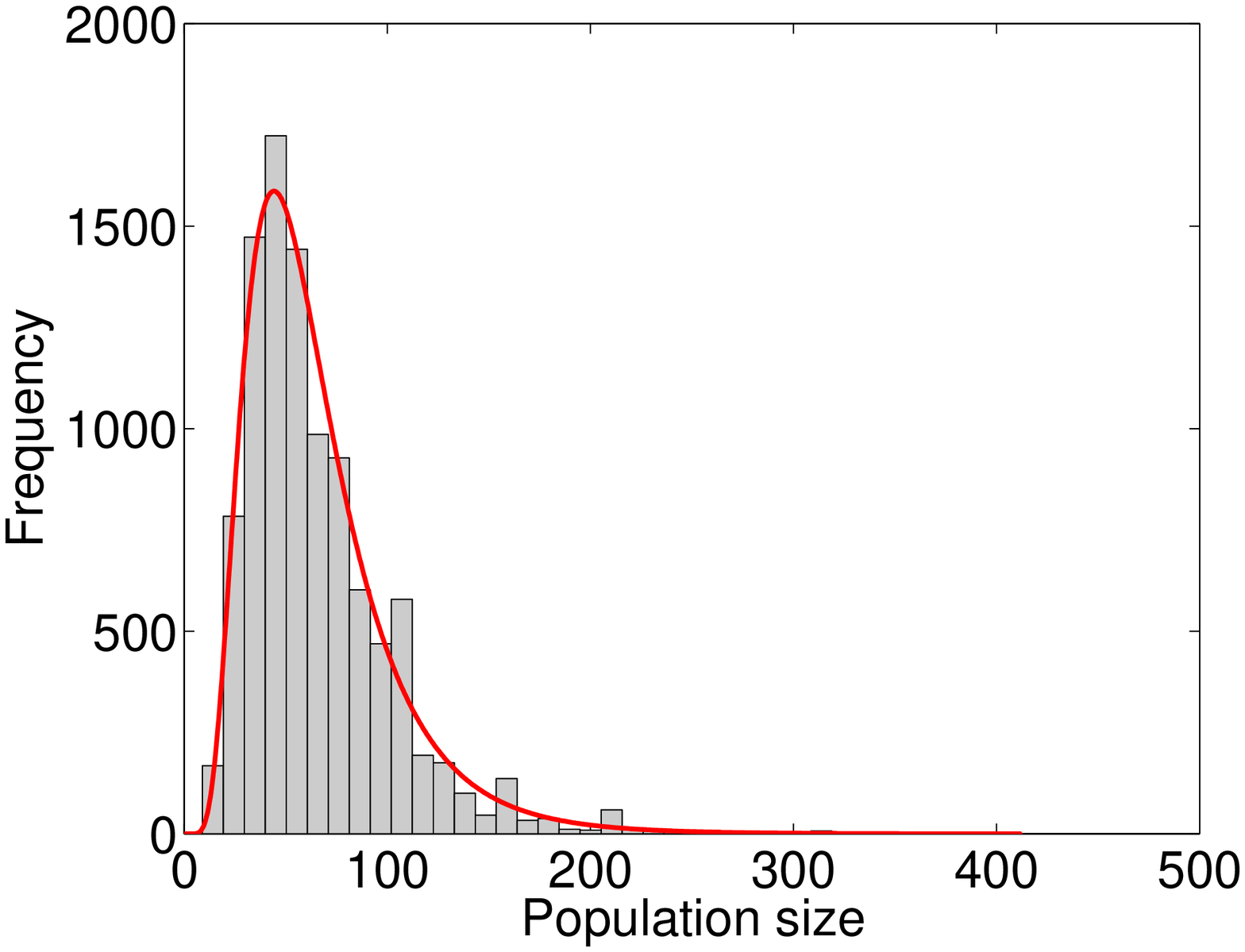,width=0.330\textwidth}
\epsfig{file=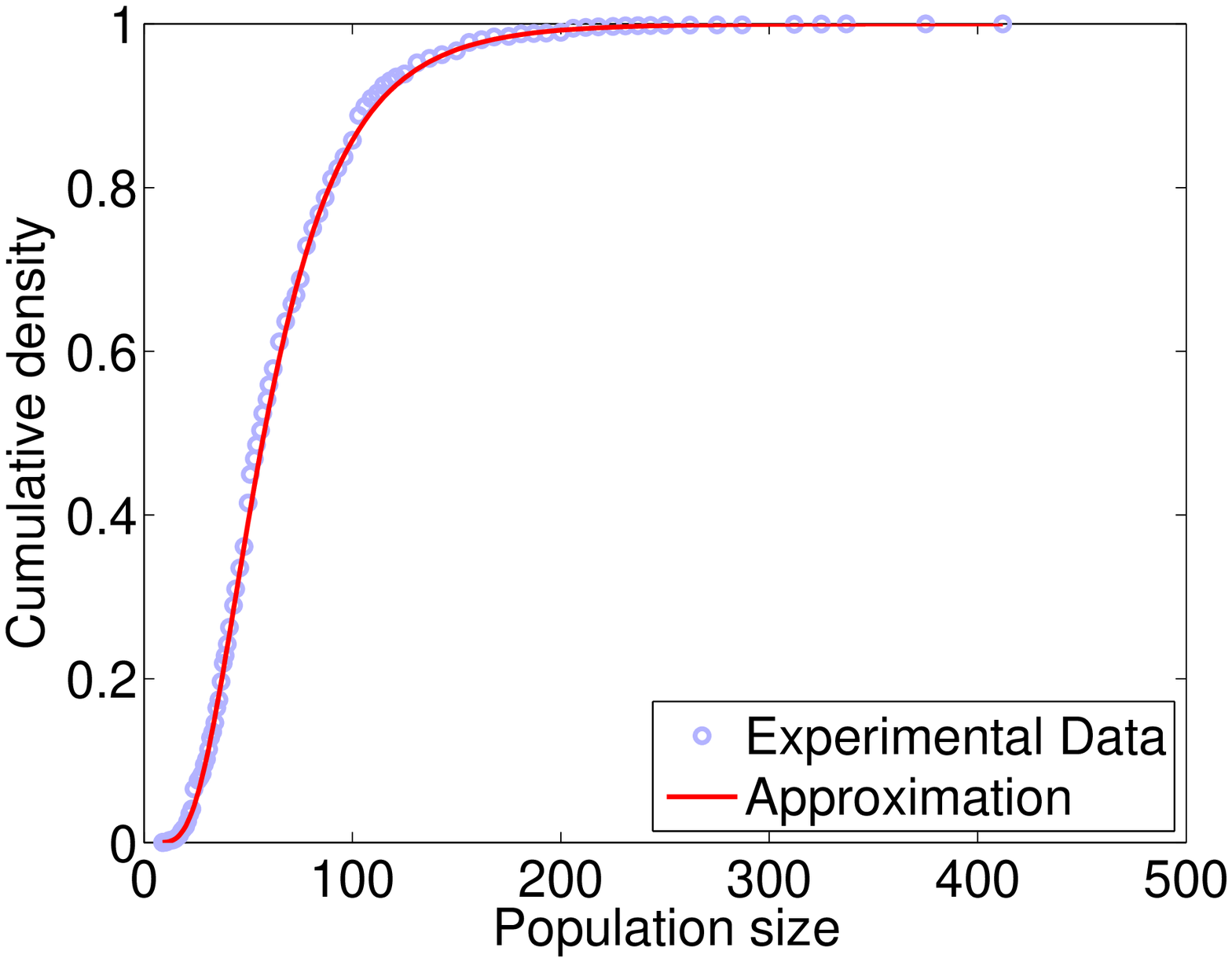,width=0.330\textwidth}
\epsfig{file=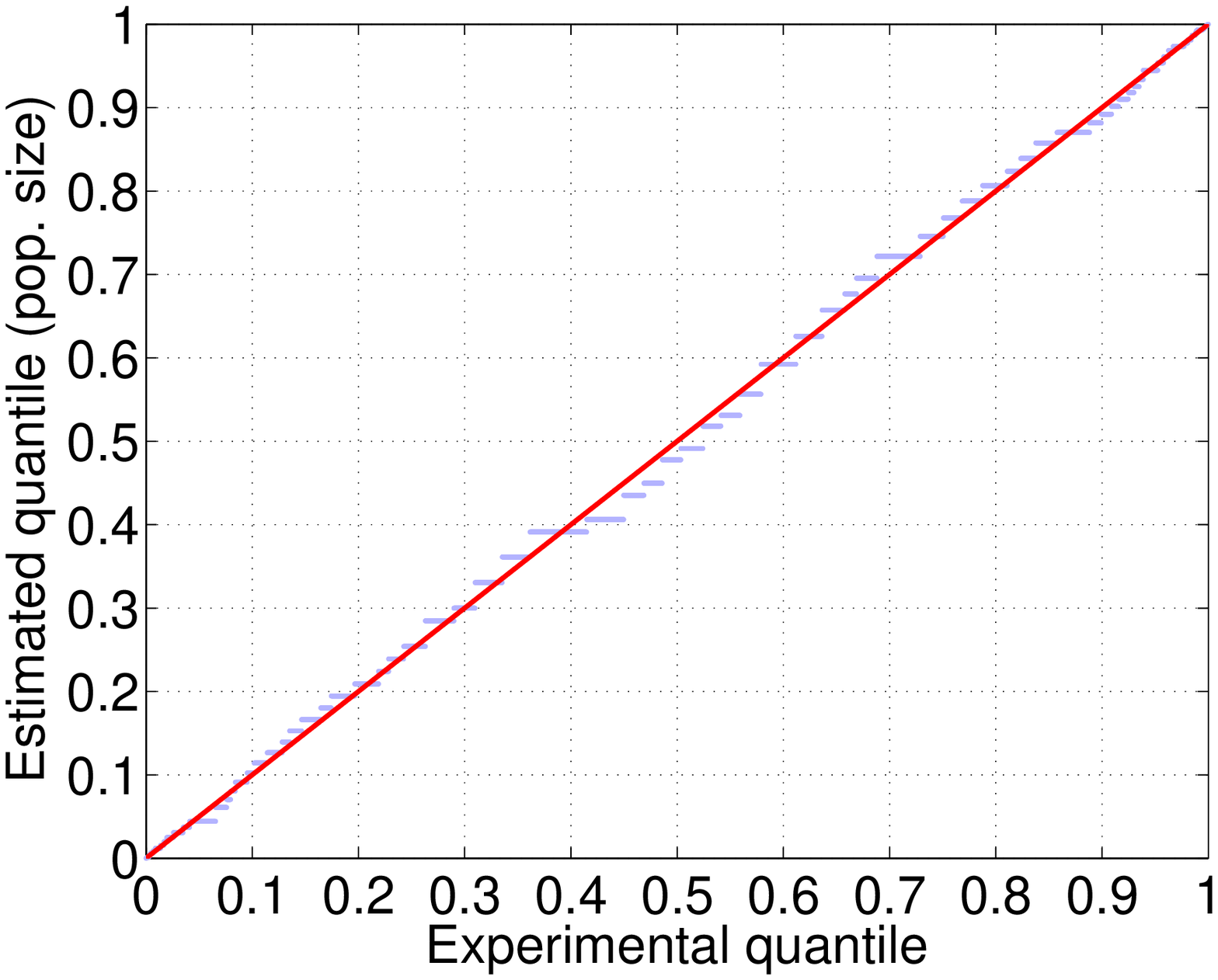,width=0.330\textwidth}
\\
\epsfig{file=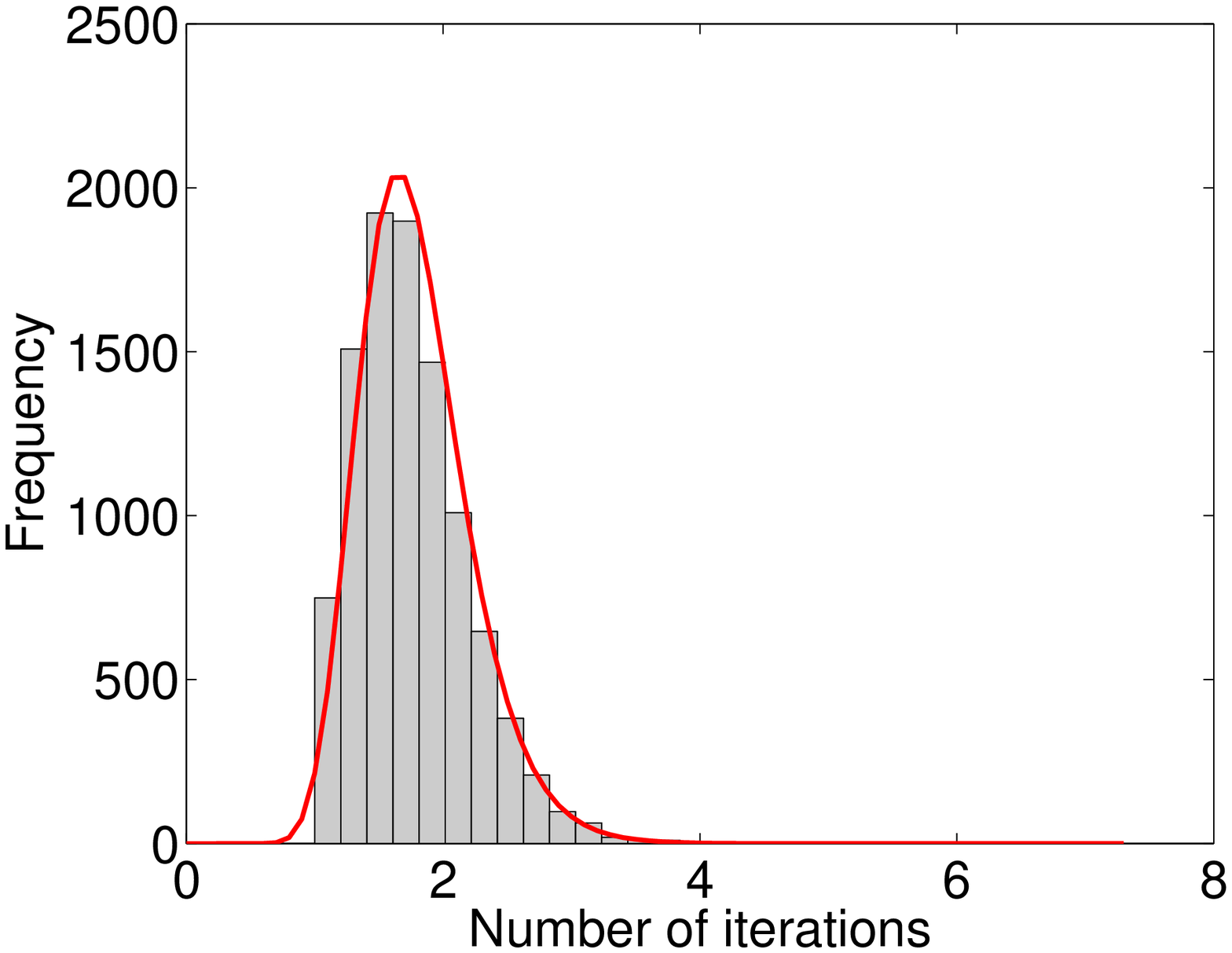,width=0.330\textwidth}
\epsfig{file=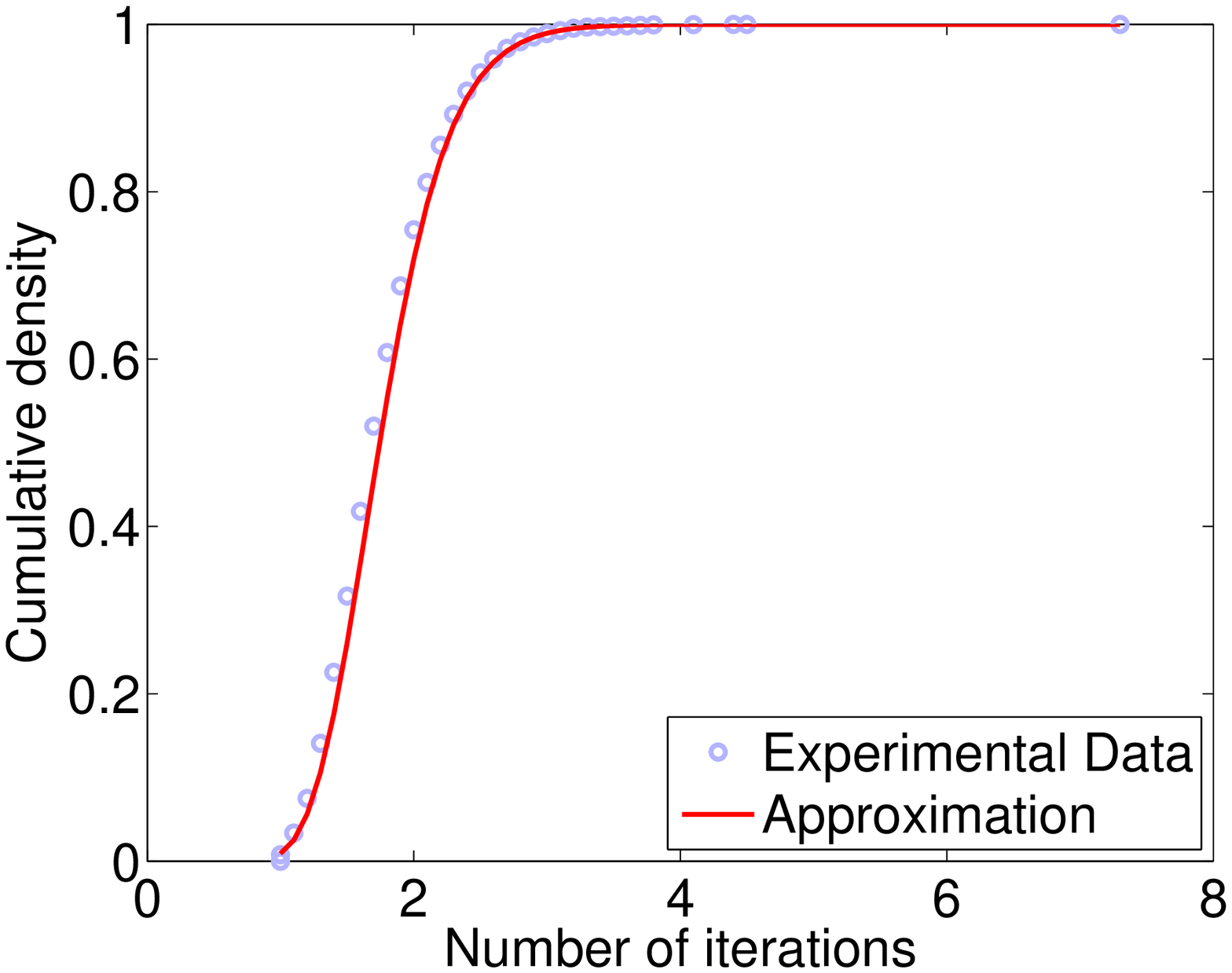,width=0.330\textwidth}
\epsfig{file=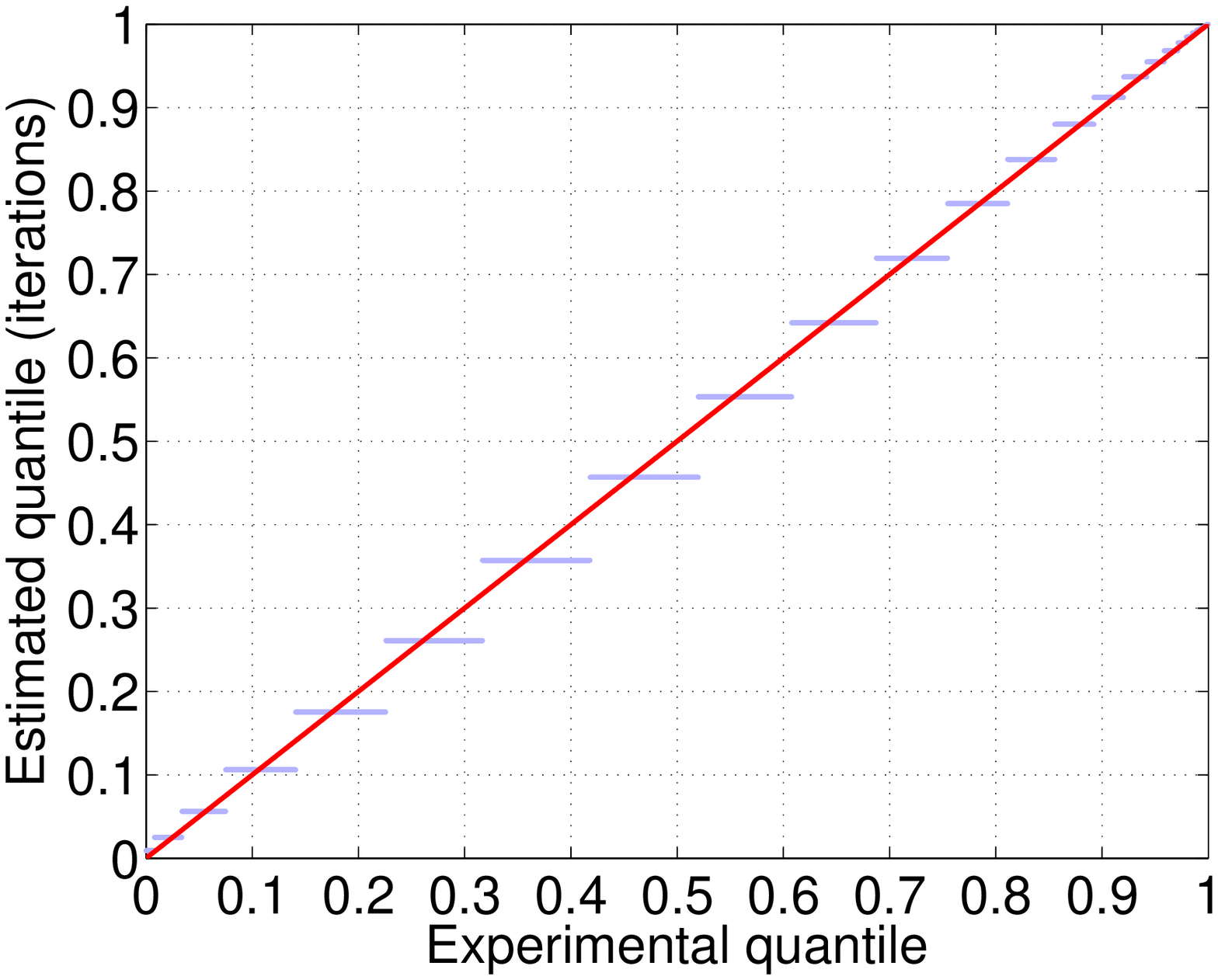,width=0.330\textwidth}
\\
\epsfig{file=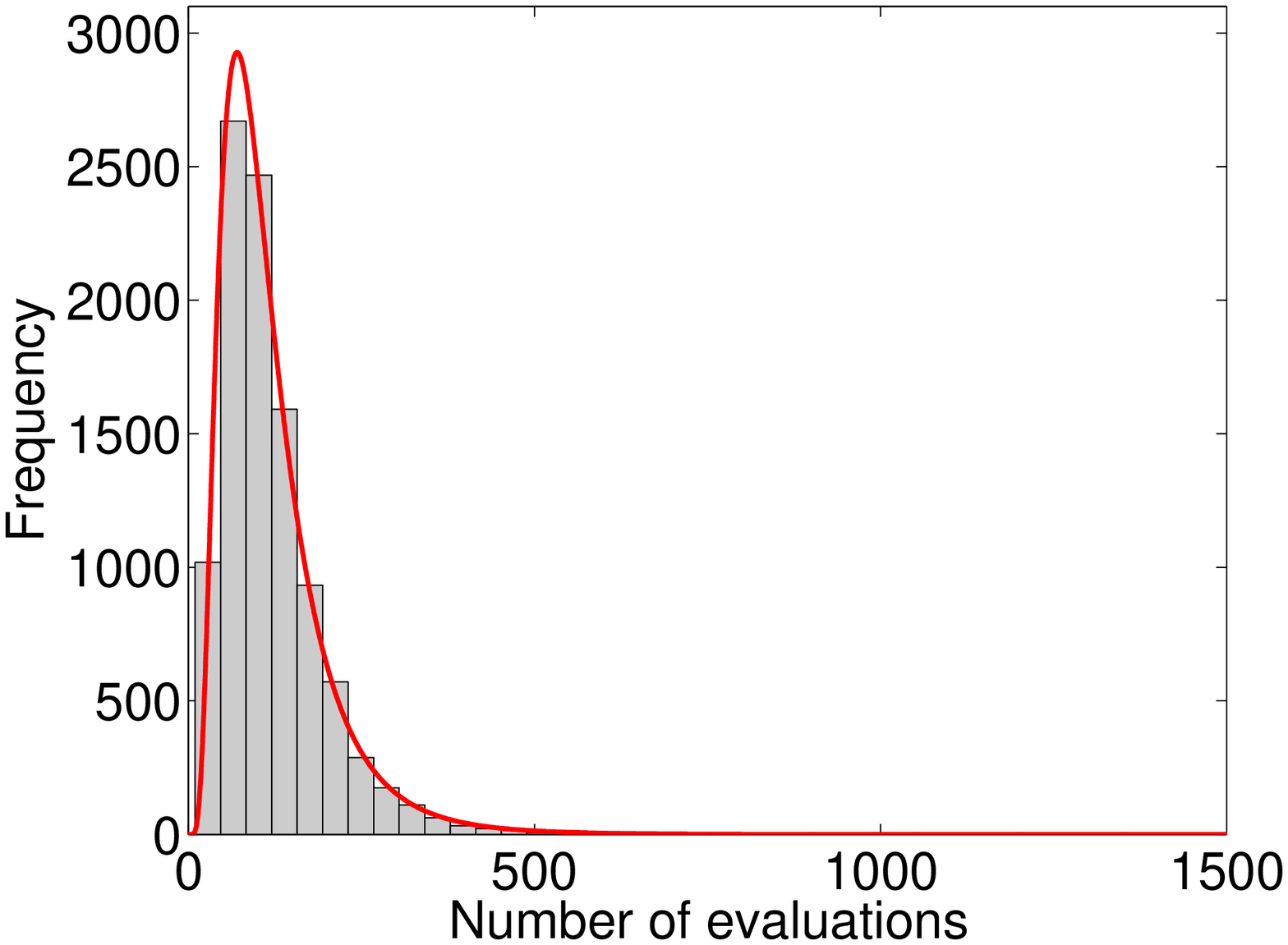,width=0.330\textwidth}
\epsfig{file=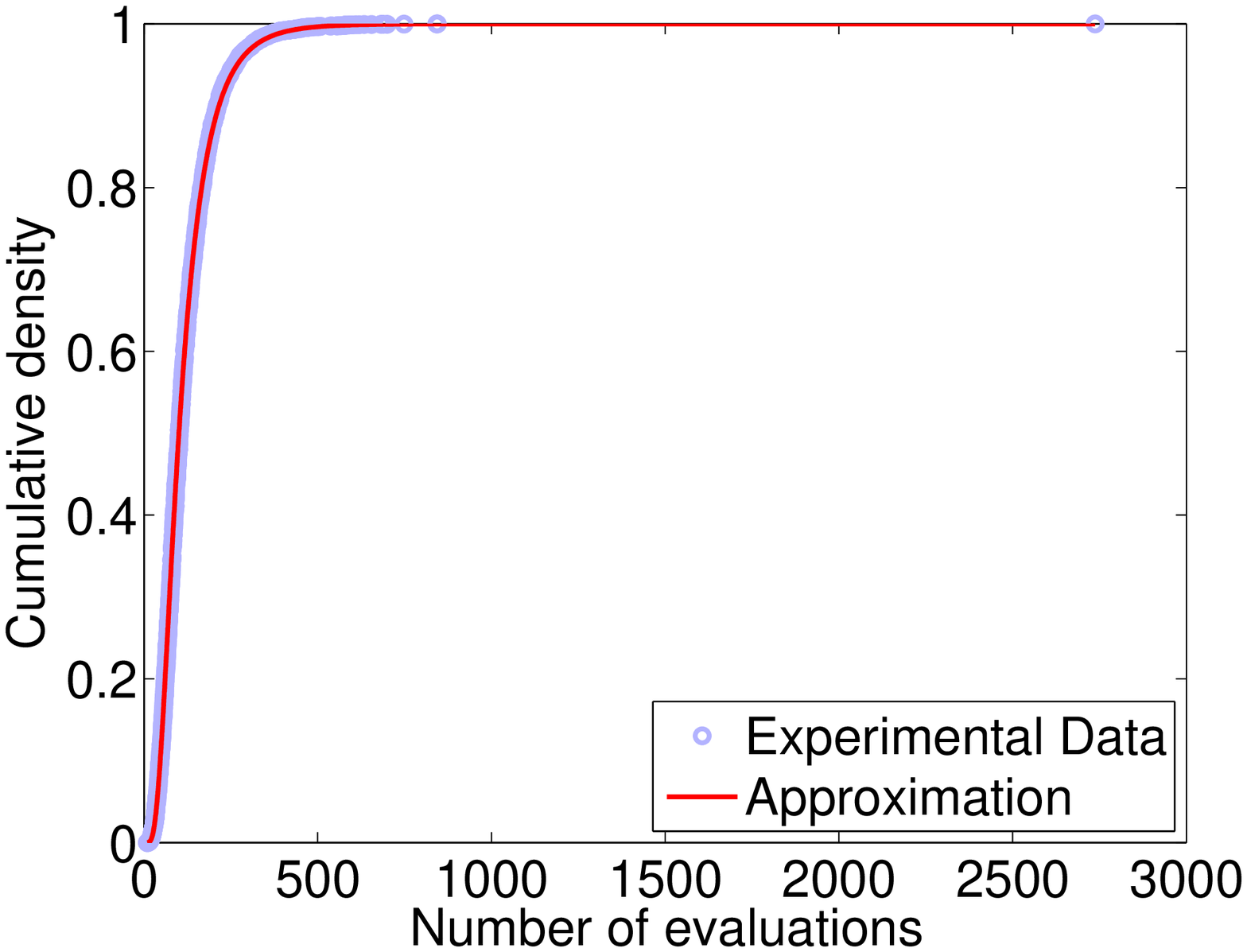,width=0.330\textwidth}
\epsfig{file=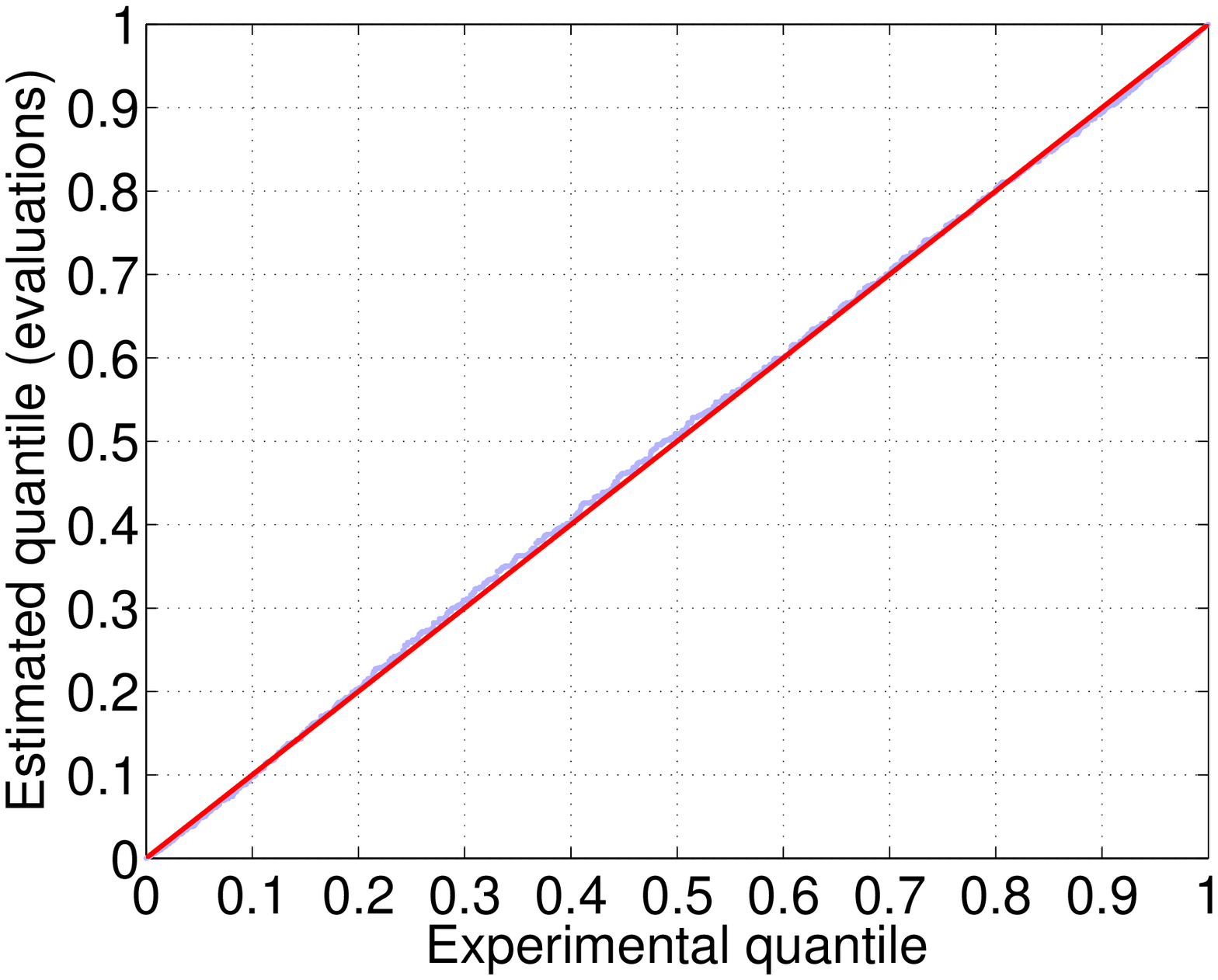,width=0.330\textwidth}
\\
\epsfig{file=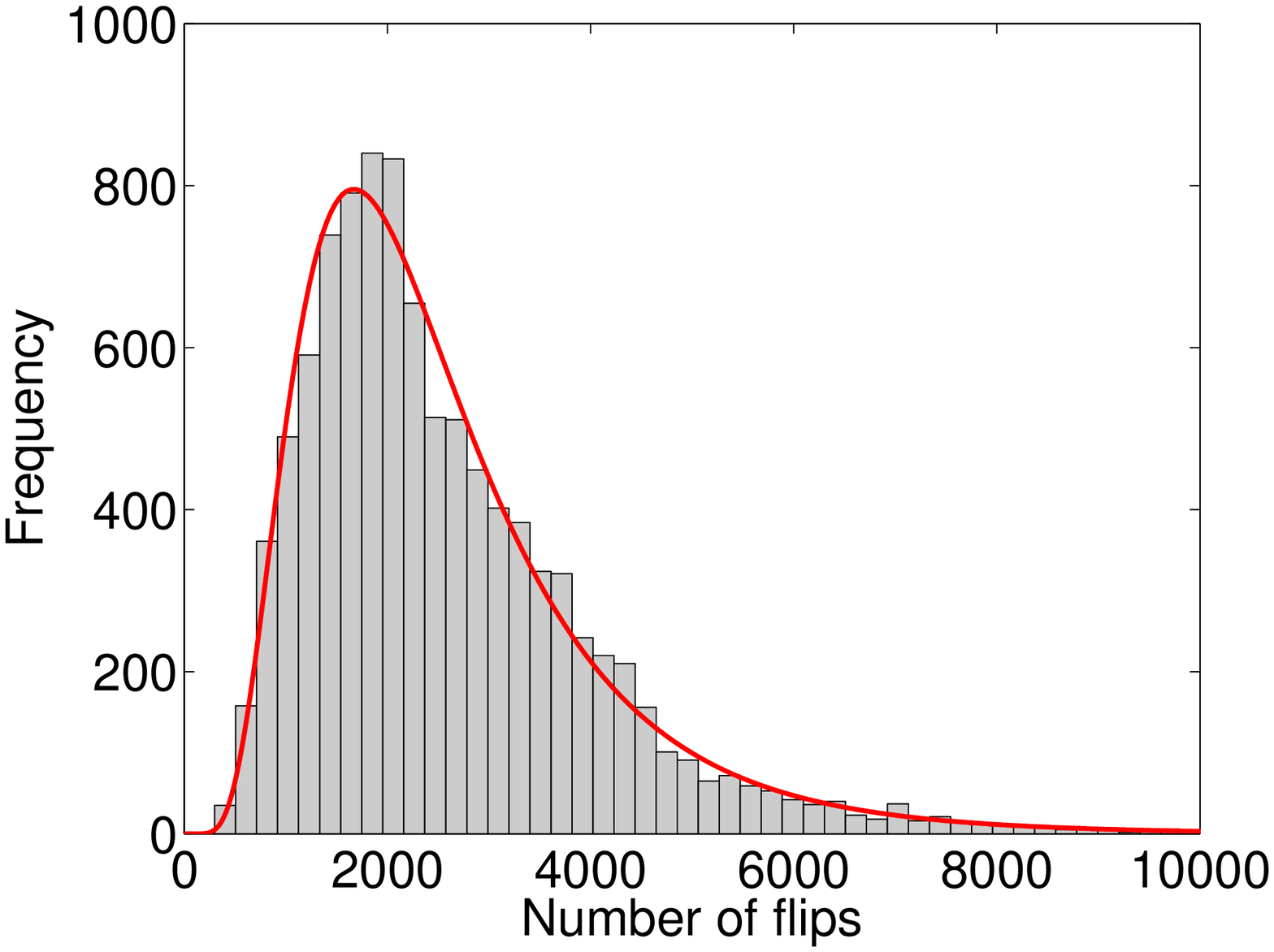,width=0.330\textwidth}
\epsfig{file=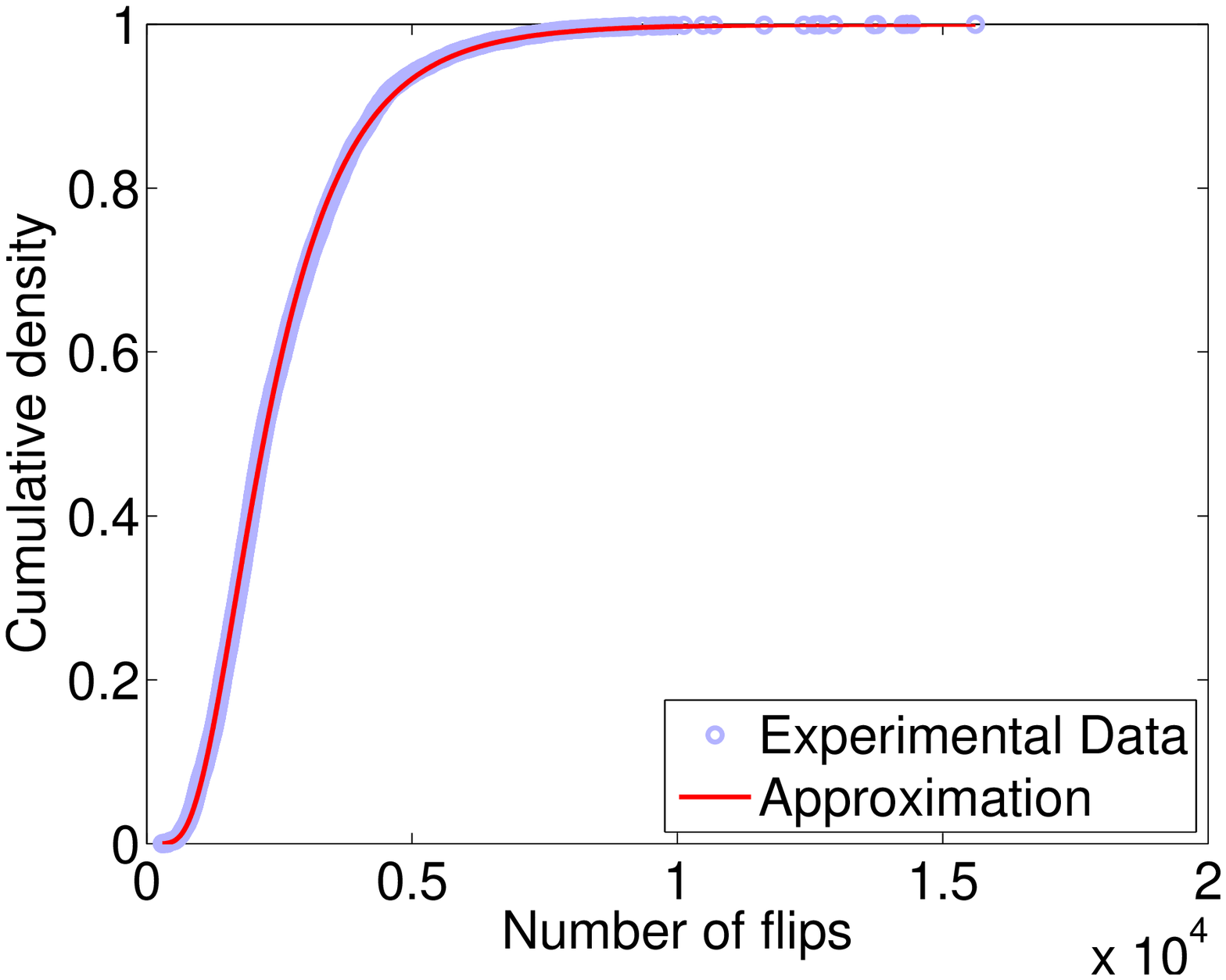,width=0.330\textwidth}
\epsfig{file=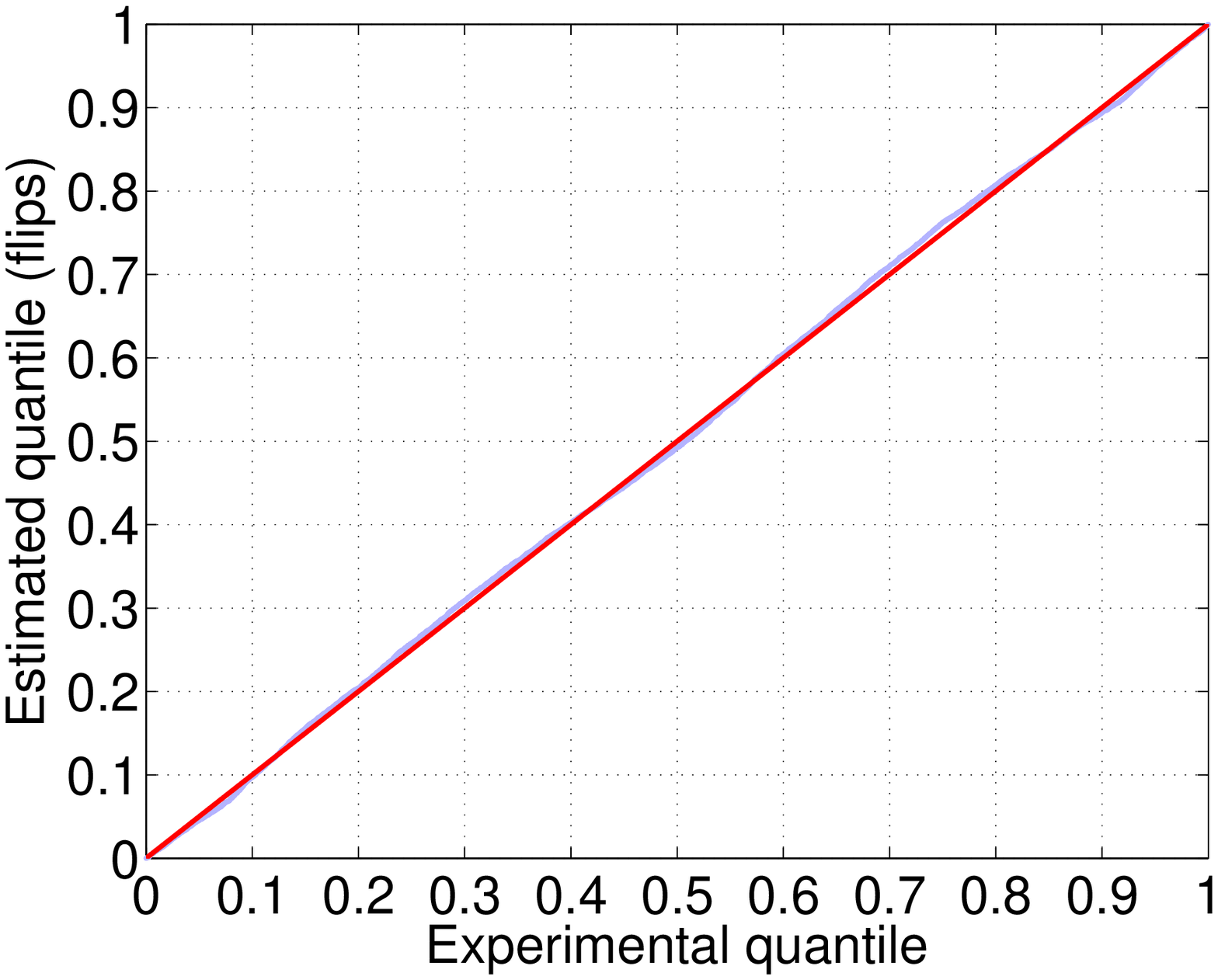,width=0.330\textwidth}
\caption{
A comparison of the experimental results (the population size, the
number of iterations, the number of evaluations, and the number
of flips---from top to bottom) and the log-normal distribution
estimates with the parameters obtained using the unbiased estimator
for $n=80$. The left-hand side shows the normalized probability
density function with the histogram from experimental data, the middle
graphs show the cumulative distribution function with the cumulative
experimental frequencies, and the right-hand size shows the Q-Q plot.}
\label{fig-distributions-small}
\end{figure}

Figure~\ref{fig-distributions-small} illustrates the match between the
estimated log-normal distribution and the experimental data for the
population size, the number of iterations, the number of evaluations,
and the number of flips for $n=80$. Analogous matches were found for
smaller problem sizes.

\begin{figure}
\hspace*{4.5ex}
{\epsfig{file=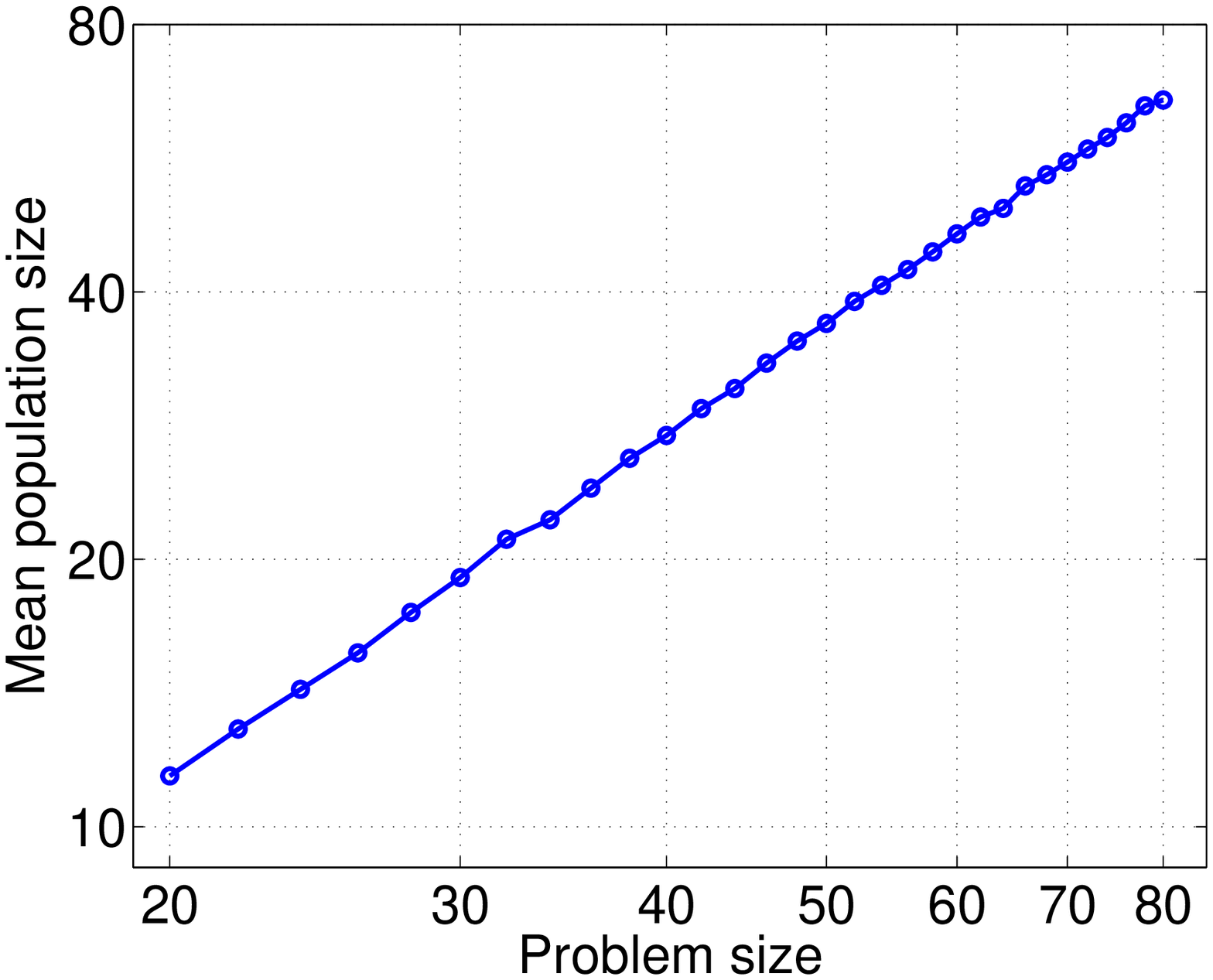,width=0.400\textwidth}}
\hspace*{4.5ex}
{\epsfig{file=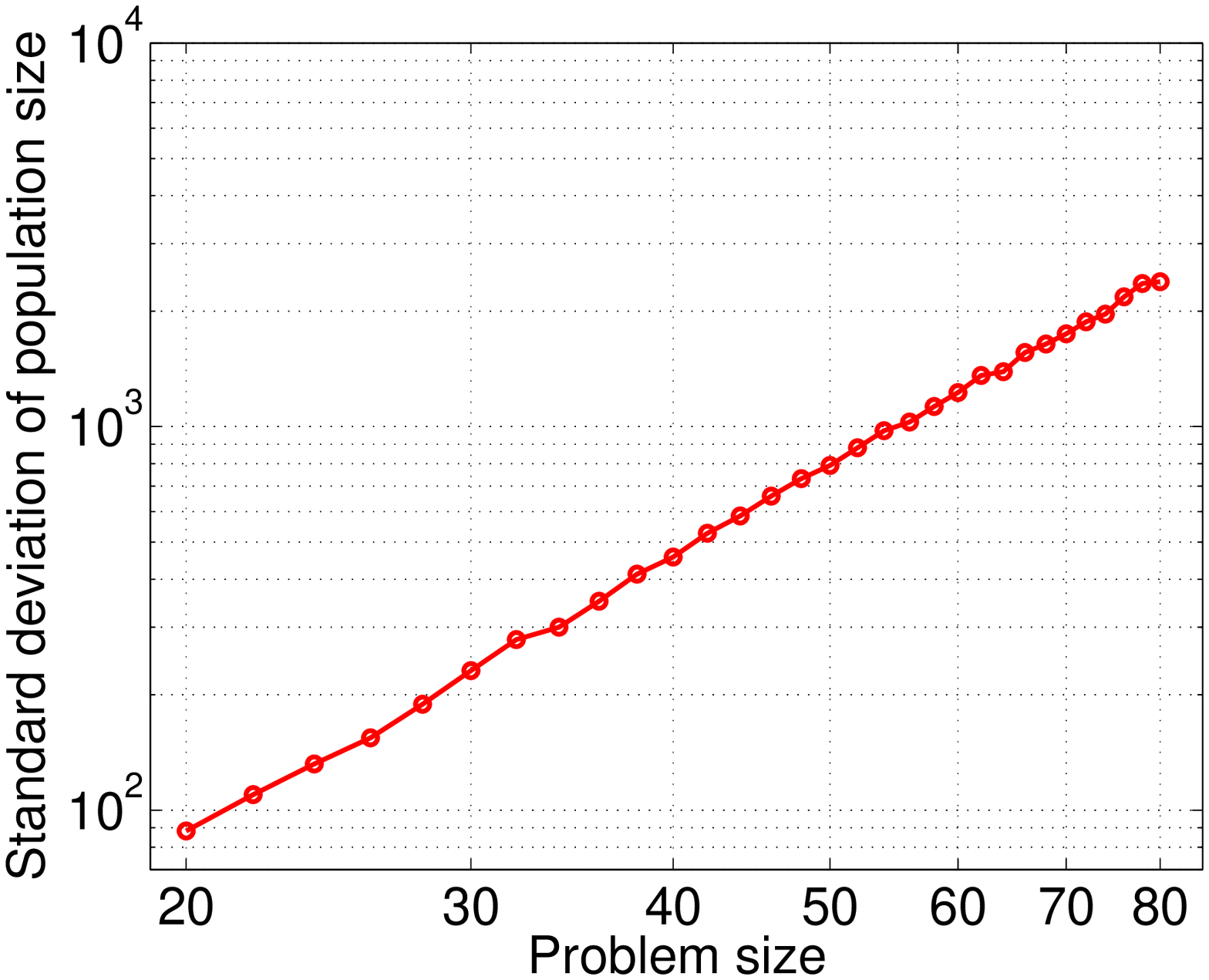,width=0.400\textwidth}}
\hspace*{4.5ex}\\
\hspace*{4.5ex}
{\epsfig{file=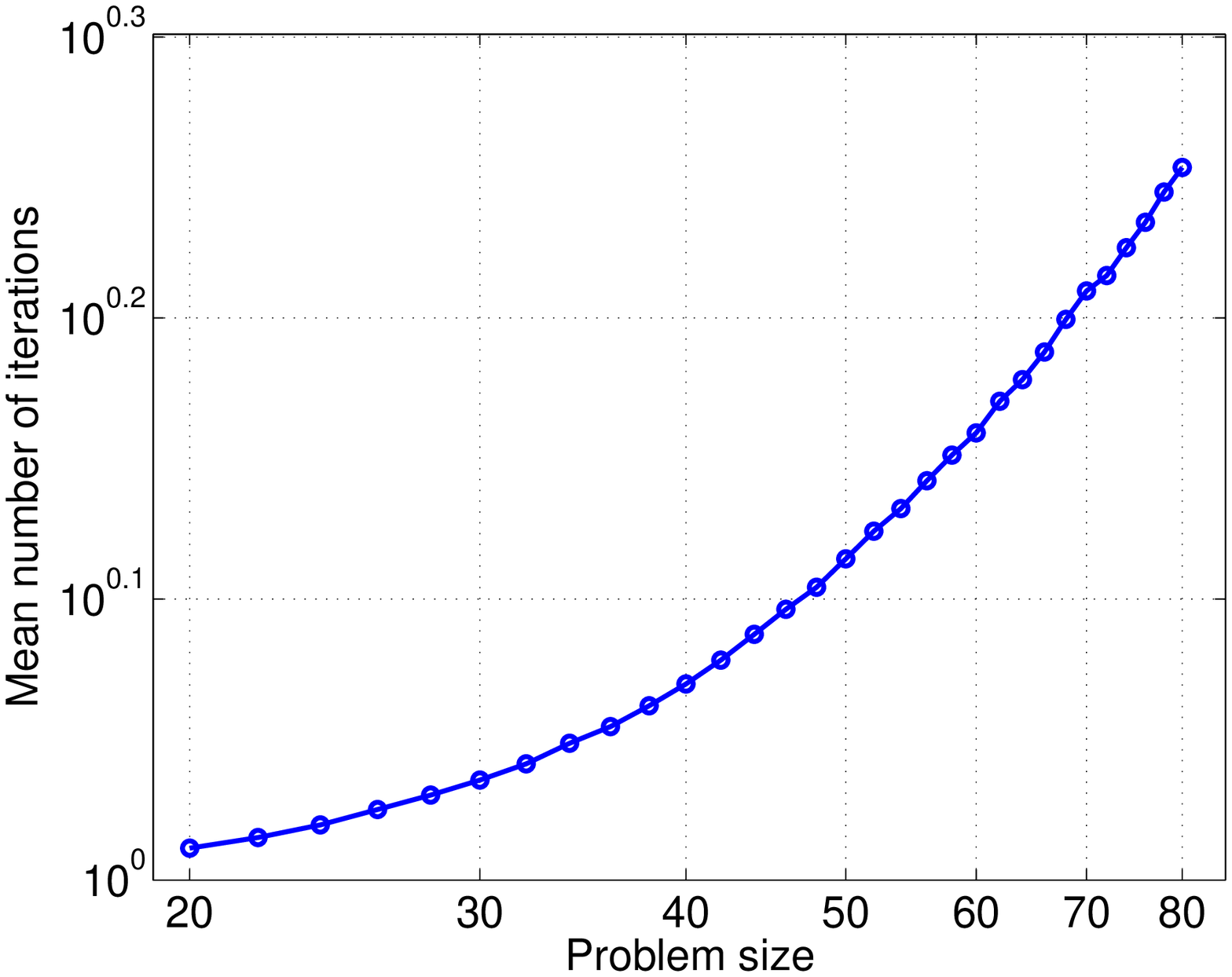,width=0.400\textwidth}}
\hspace*{4.5ex}
{\epsfig{file=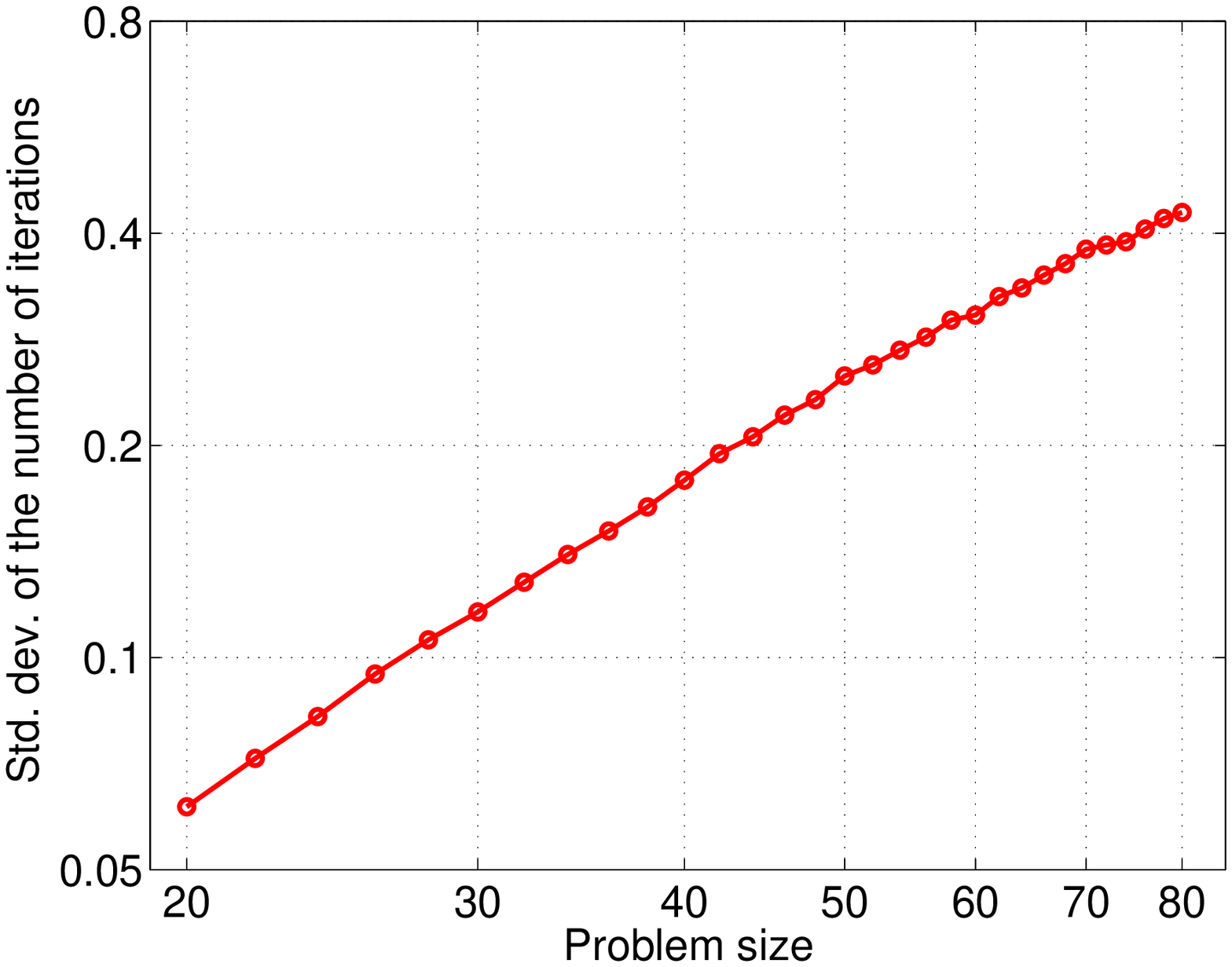,width=0.400\textwidth}}
\hspace*{4.5ex}\\
\hspace*{4.5ex}
{\epsfig{file=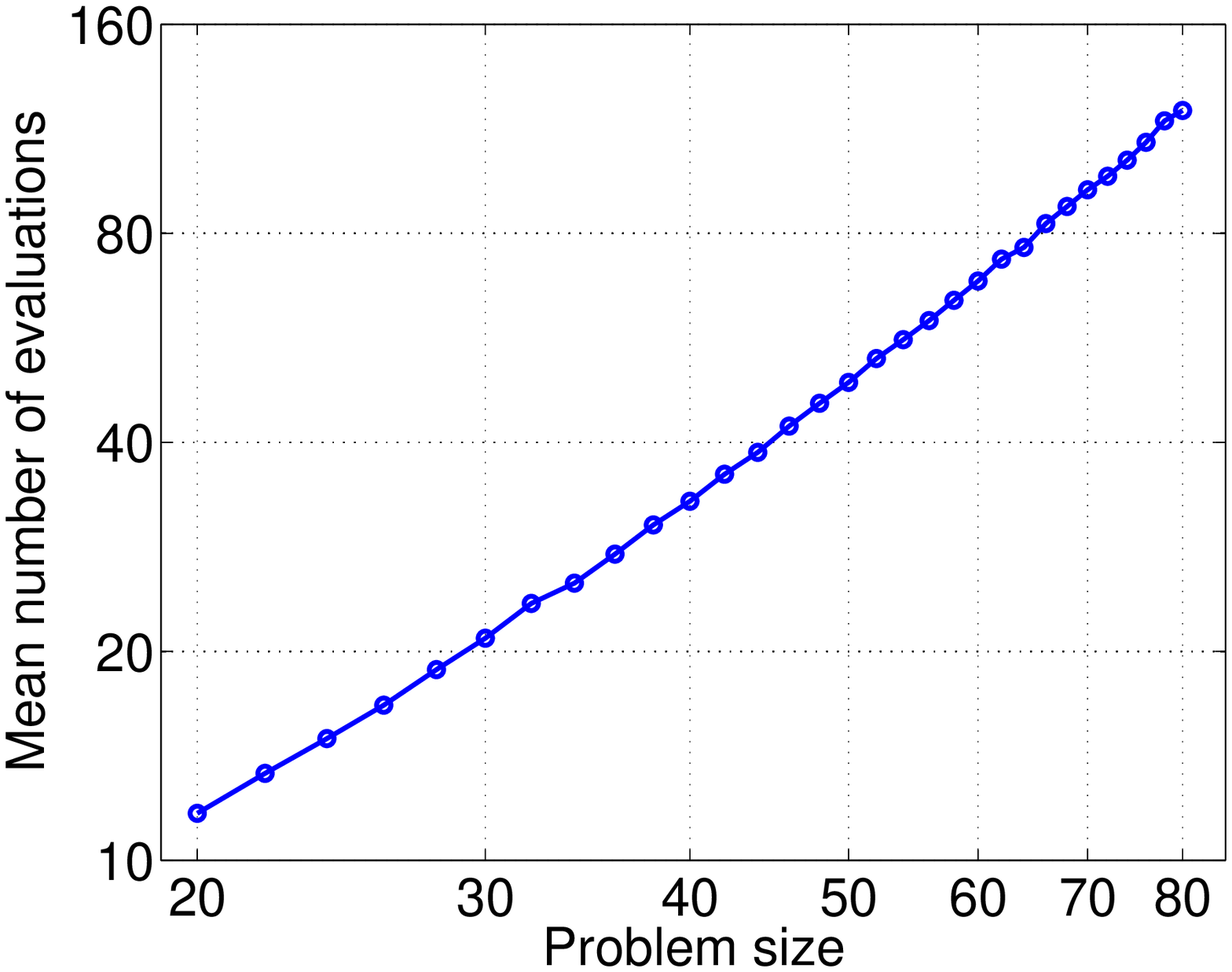,width=0.400\textwidth}}
\hspace*{4.5ex}
{\epsfig{file=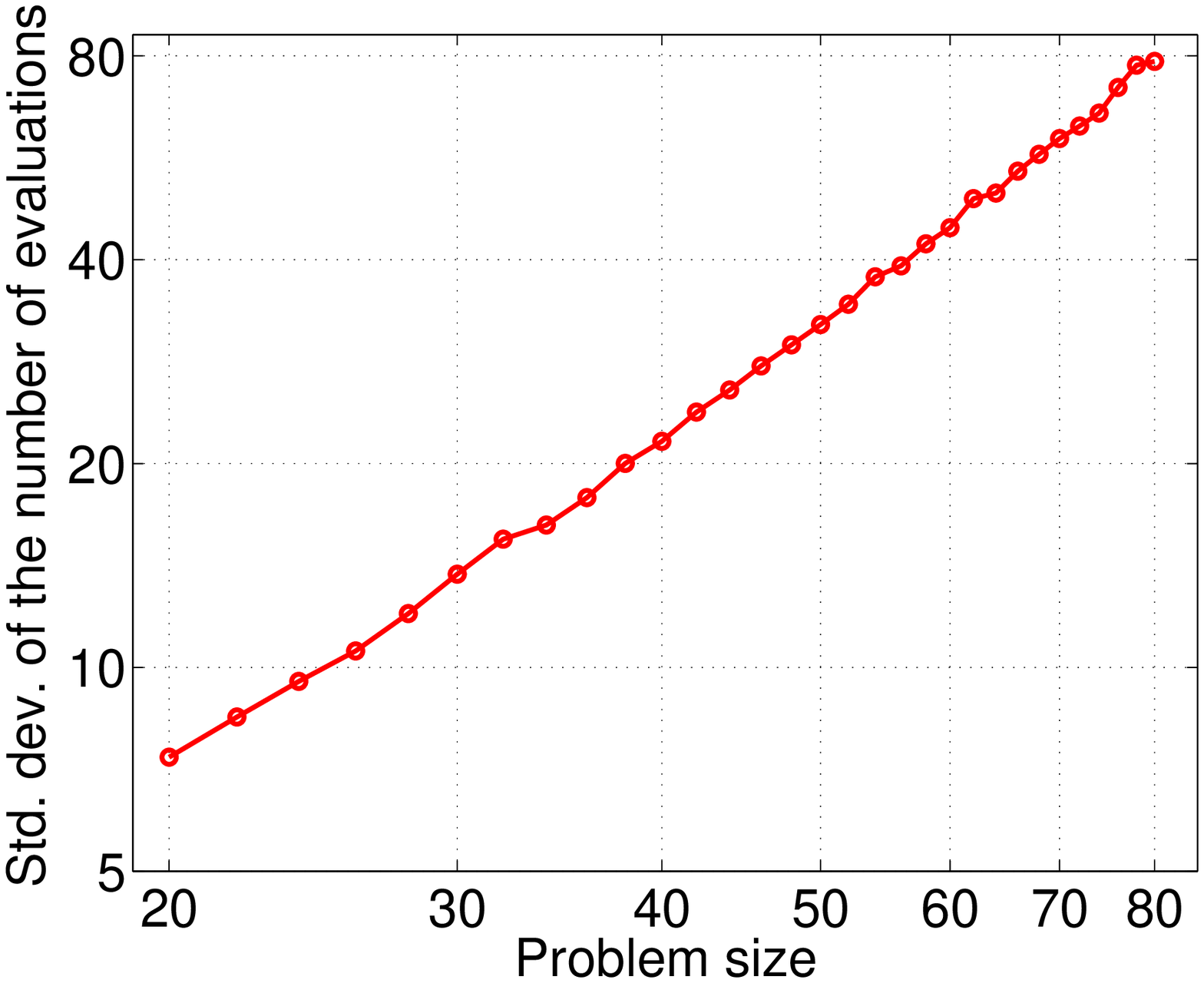,width=0.400\textwidth}}
\hspace*{4.5ex}\\
\hspace*{4.5ex}
{\epsfig{file=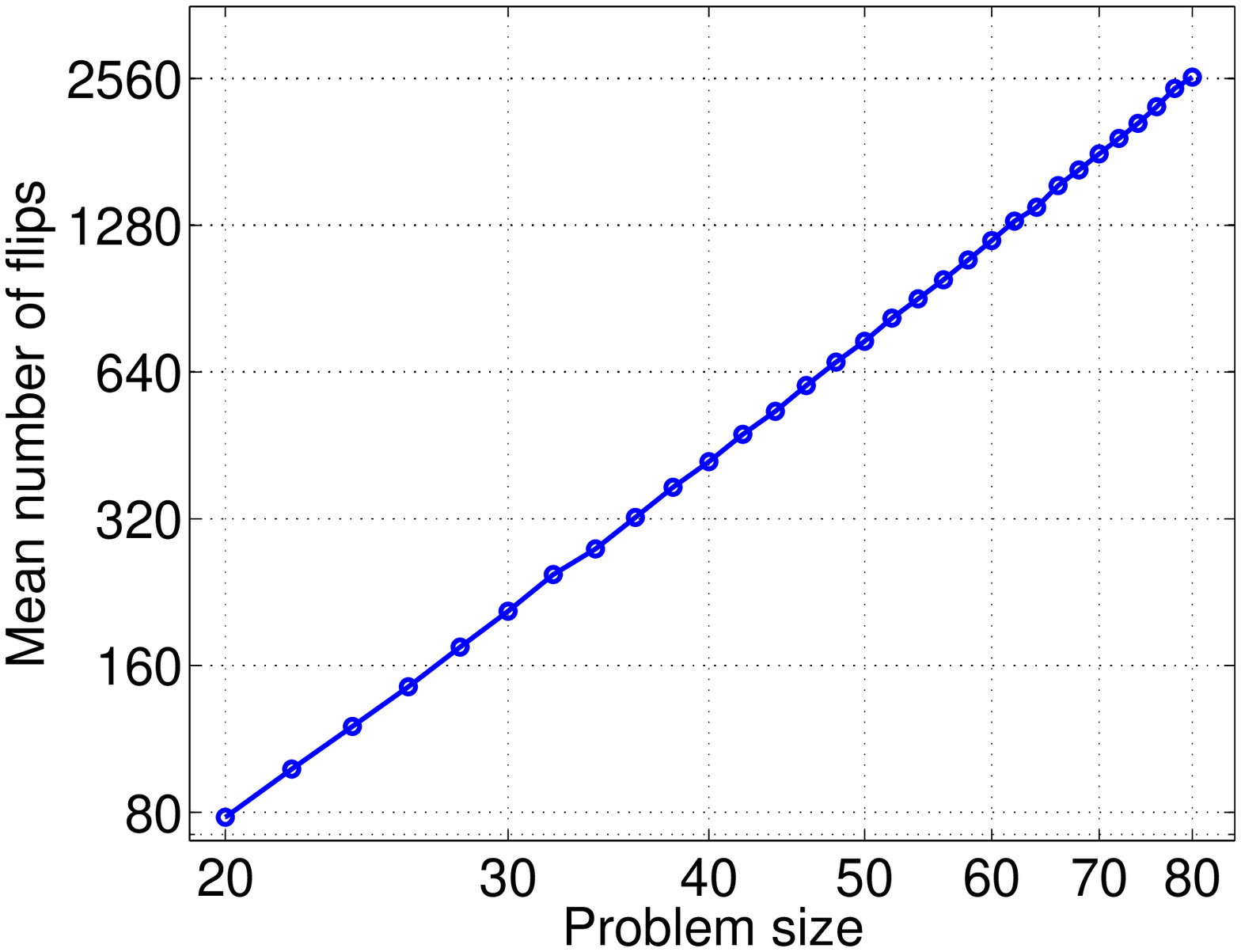,width=0.400\textwidth}}
\hspace*{4.5ex}
{\epsfig{file=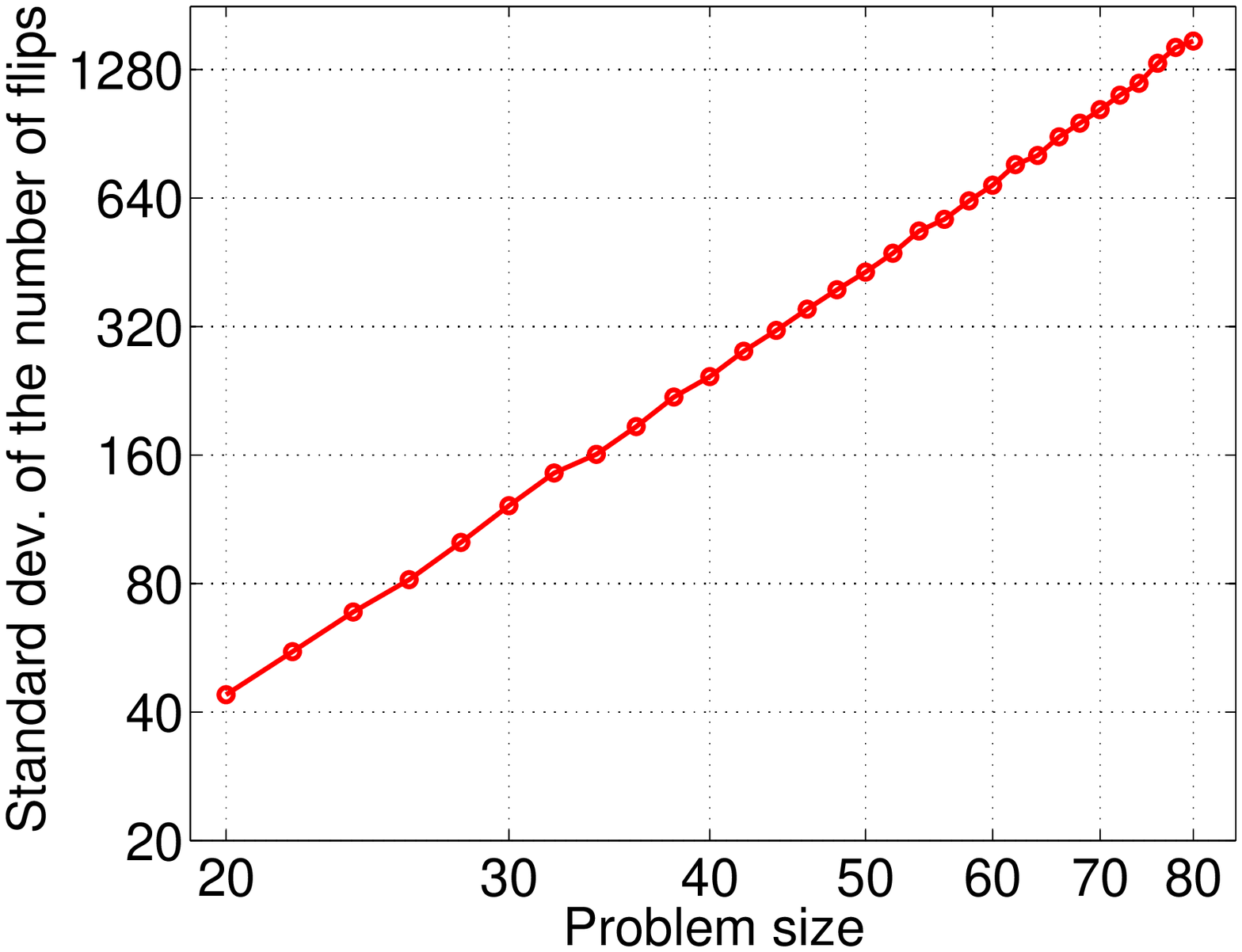,width=0.400\textwidth}}
\hspace*{4.5ex}
\caption{
Mean and standard deviation of the log-normal approximation for the
population size, the number of iterations, the number of evaluations,
and the number of flips for SK spin glasses of $n=20$ to $n=80$
in with step $2$.}
\label{fig-mean-stdev-small}
\end{figure}

Figure~\ref{fig-mean-stdev-small} shows the mean and standard deviation
of the distribution estimates for the entire range of SK instances
from $n=20$ to $n=80$ with step 2. The results indicate that while the
population size appears to grow only polynomially fast, the remaining
quantities appear to grow exponentially fast, which is mirrored by the
estimates of the number of iterations. Nonetheless, for the number of
iterations, a significant factor influencing the rate of growth is that
the average number of iterations for small problems is close to $1$ and
the number of iterations has to be always at least $1$; therefore, only
the results for larger problems will provide a clear indication of how
fast the number of iterations actually grows on average. Furthermore,
it is important to not only study the mean statistics, but also to
analyze the tails of the estimated distributions. This is especially
necessary for problems like mean-field SK spin glasses for which the
difficulty of problem instances varies significantly and, as a result,
while many instances are relatively easy, solving the most difficult
instances becomes especially challenging.

\section{How to Locate Global Optima Reliably for Bigger Problems?}
\label{section-how-to-bigger}

To make sure that hBOA finds a ground state reliably, it is necessary
to set the population size and the maximum number of iterations to
sufficiently large values. The larger the population size and the
number of iterations, the more likely hBOA finds the optimum. Of
course, as the problem size increases, the population-sizing and
time-to-convergence requirements will increase as well, just like
was indicated by the initial results presented in the previous section.

In this section we present three approaches to reliably locate
the ground states of SK spin-glass instances unsolvable with the
branch-and-bound algorithm. The first two approaches are based on the
statistical models of the population size and the number of iterations
for smaller problem instances, such as those developed in the previous
section. On the other hand, the last approach does not require any
statistical model or prior experiments, although it still requires an
estimate of the upper bound on the maximum number of iterations. The
proposed approaches are not limited to the SK spin-glass model and
can thus be used to reliably identify the global optima of other
difficult problems.

\subsection{Modeling the Percentiles}

The first approach is based on modeling the growth of the percentiles
of the estimated probability distributions. As the input, we use
the estimated distributions of the population size and the number of
iterations for spin-glass instances of sizes $n\leq 80$.

For each problem size $n$, we first compute the $99.999$ percentile
of the population-size model so that it is ensured that the resulting
population sizes will be sufficiently large for all but the $0.001\%$
most difficult instances. Then, we approximate the growth of the
$99.999$ percentile and use this approximation to predict sufficient
population sizes for larger problems. Since the estimation of
the growth function is also subject to error, we can use a $95\%$
confidence bound for the new predictions and choose the upper bound
given by this confidence bound. An analogous approach can then be
used for the number of iterations.

Of course, the two confidence bounds involved in this approach can
be changed and the estimation should be applicable regardless of
the model used to fit the distribution of the population size and
the number of iterations; nonetheless, it is important to build an
accurate approximation of the true distribution of the population
sizes in order to have an accurate enough approximation of the chosen
percentile. Furthermore, it is important to use an appropriate growth
function to predict the percentiles in order to minimize the errors
for predicting the parameters for bigger problems.

The same approach can be used to predict the population size and the
number of iterations for any population-based evolutionary algorithm,
such as the genetic algorithm and evolution strategies. Furthermore,
when applied to other types of stochastic optimization algorithms,
other parameters may be modeled accordingly. For example, if we were
using simulated annealing, we could model the rate of the temperature
decrease.

\begin{figure}
\hspace*{1ex}
{\epsfig{file=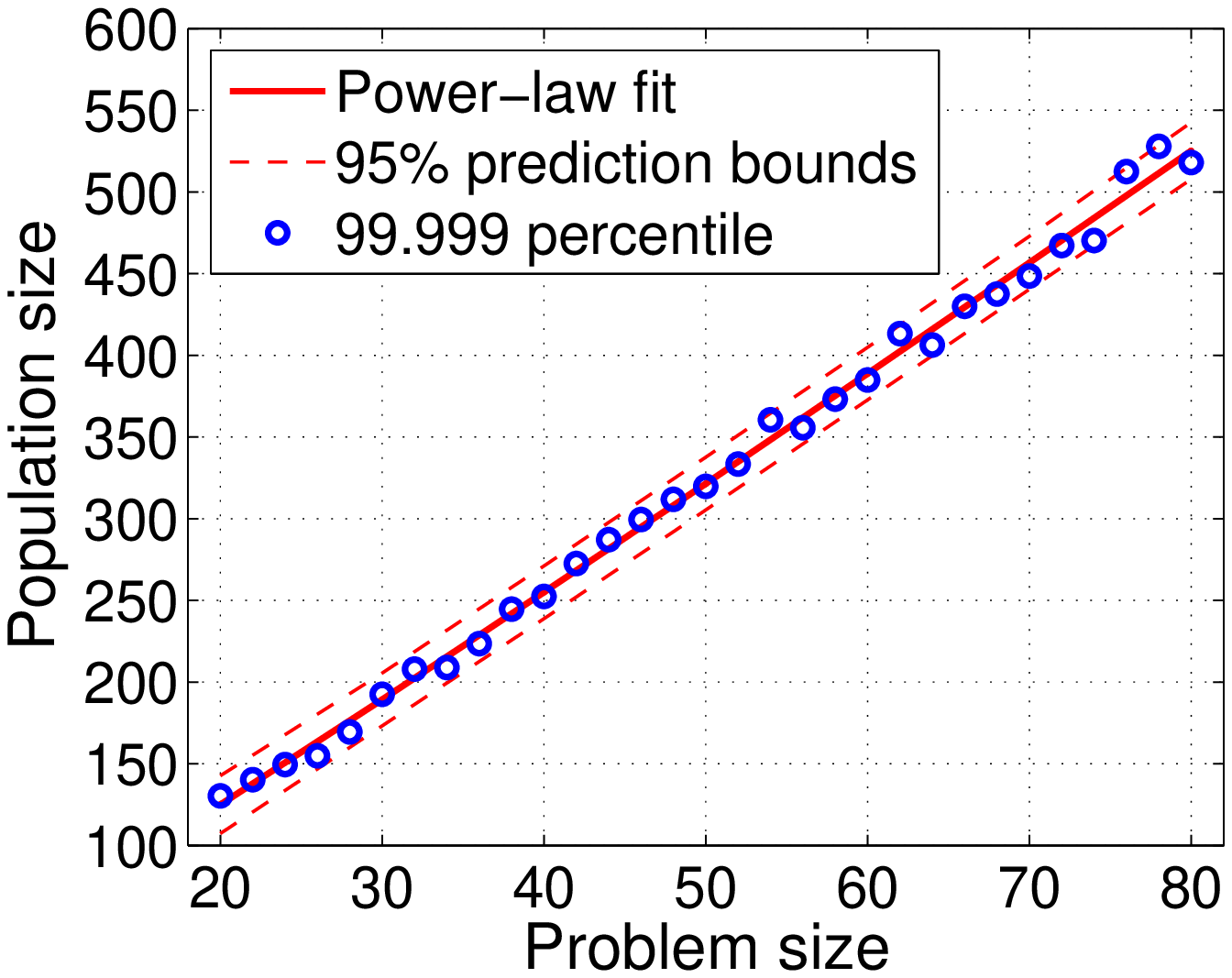,width=0.450\textwidth}}
\hspace*{3ex}
{\epsfig{file=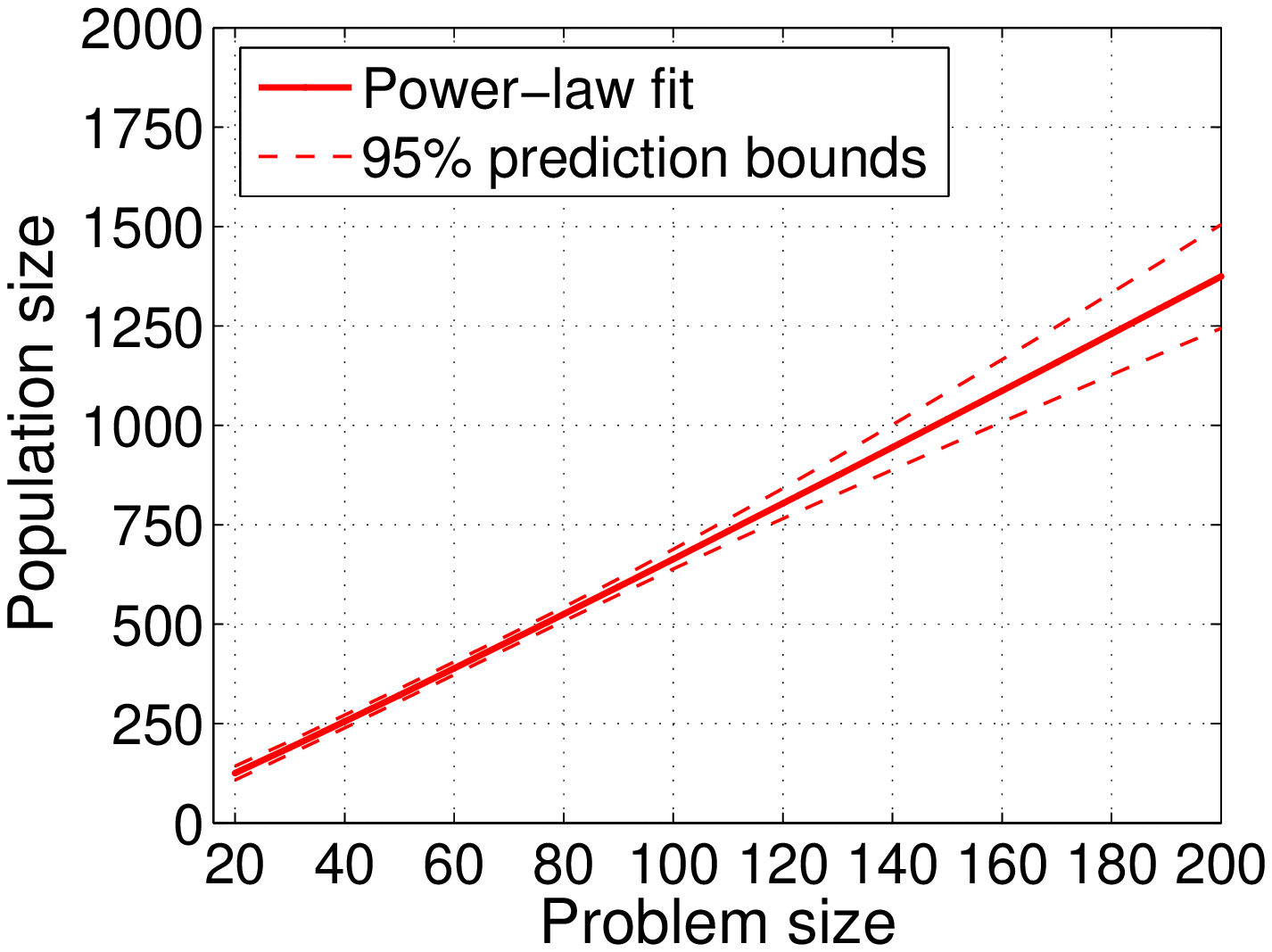,width=0.450\textwidth}}
\\
\hspace*{1ex}
{\epsfig{file=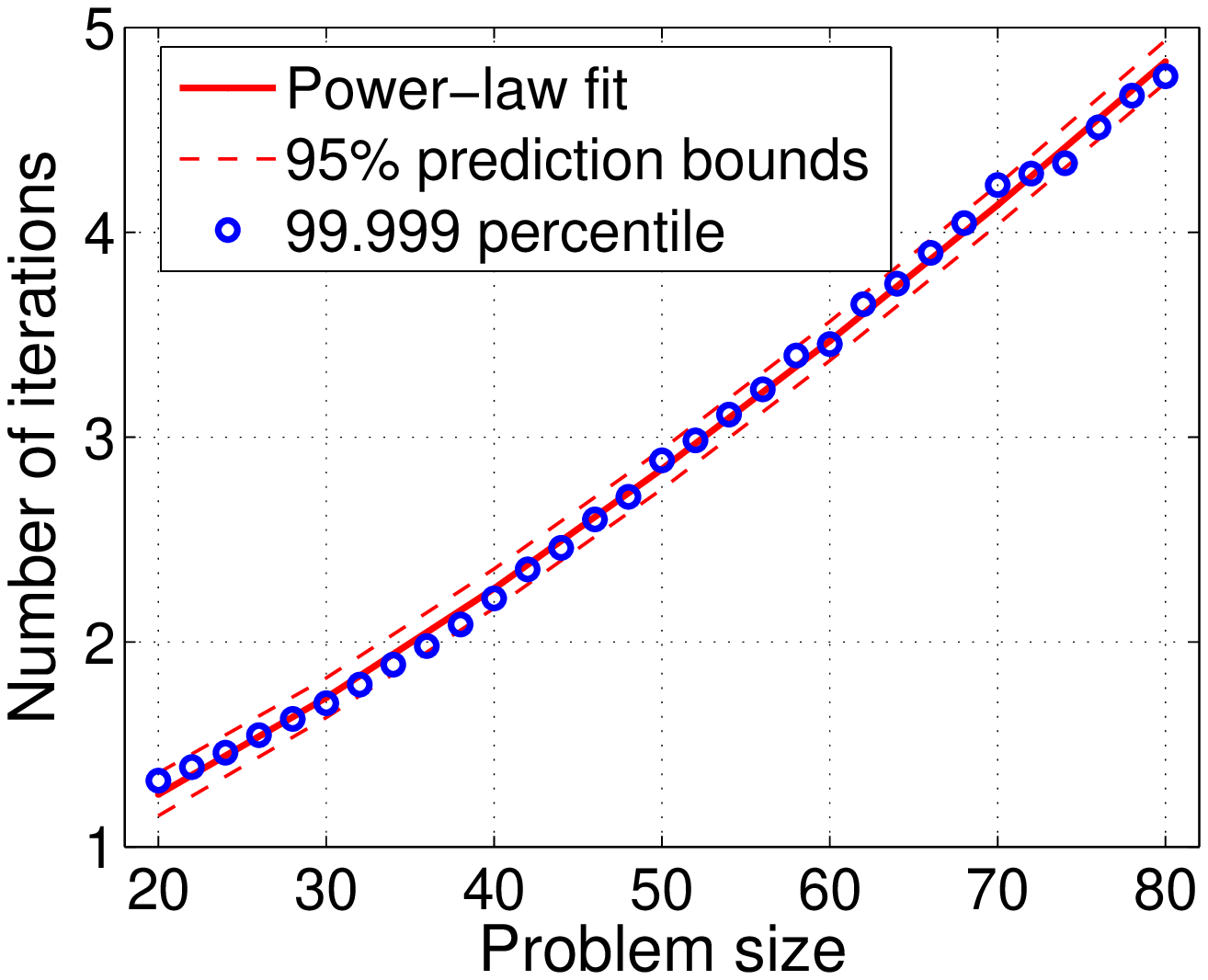,width=0.470\textwidth}}
\hspace*{2.3ex}
{\epsfig{file=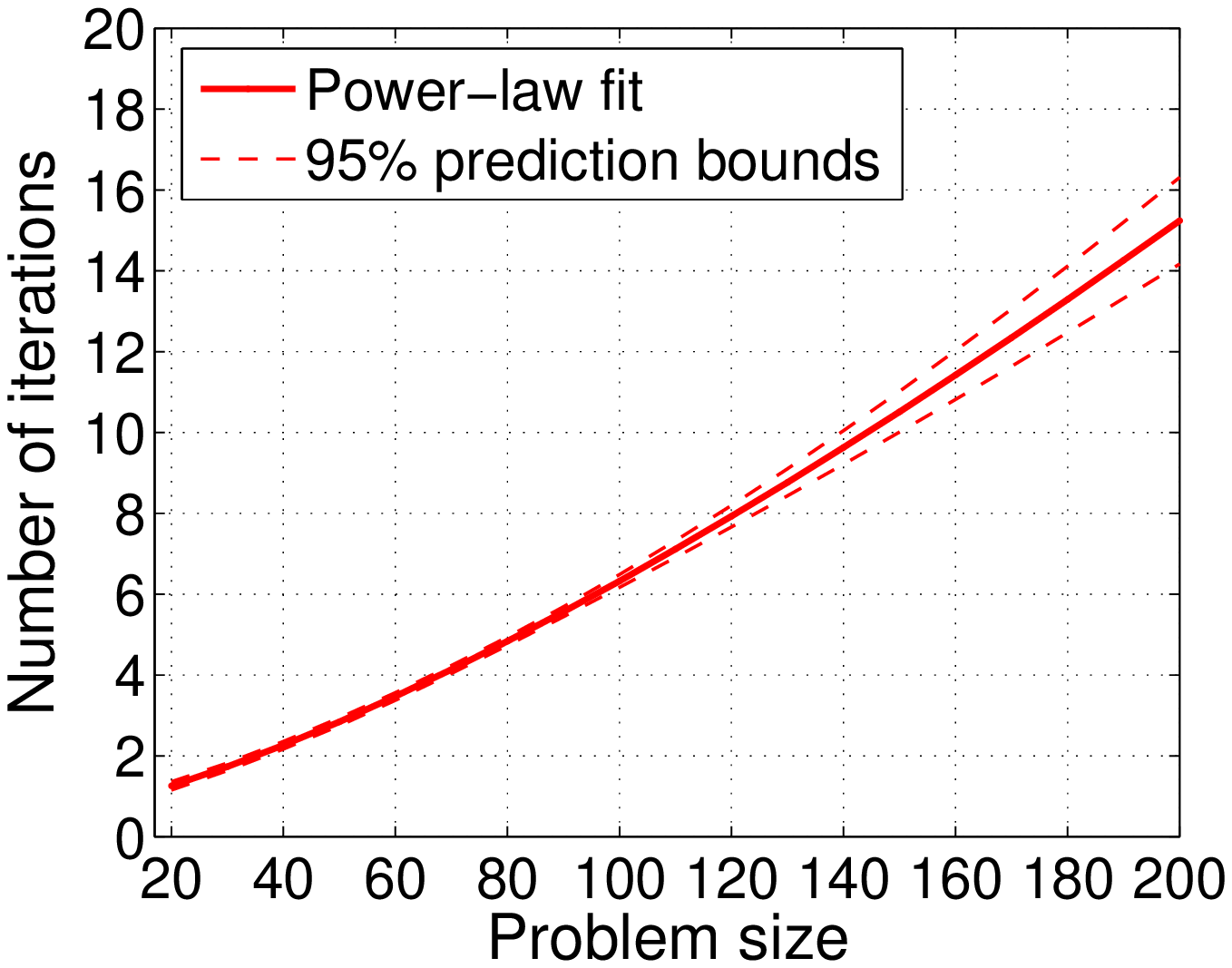,width=0.470\textwidth}}
\caption{
A model of the population size and the number of iterations based on
the growth of the 99.999 percentile of the estimated distributions. The
left-hand side shows the percentiles and the power-law fit which
estimates their growth for $n\leq 80$. The right-hand side shows the
resulting predictions for larger problems of up to $n=200$.}
\label{fig-modeling-percentiles-80}
\end{figure}

The left-hand side of figure~\ref{fig-modeling-percentiles-80} shows
the percentiles for the population size and the number of iterations
obtained from the log-normal distribution estimates for $n\leq 80$
presented in the previous section, and the best-fit curves estimating
the growth of these percentiles created with the Matlab curve-fitting
toolbox. Best approximation of the growth of both the quantities
is obtained with a power-law fit of the form $a n^b+c$ where $n$
is the number of bits (spins). The estimated parameters of the best
power-law fit of the population-size percentiles are shown below
(adjusted $R^2$ of the estimate is $0.9963$):

\vspace*{0.5em}

\begin{tabular}{|c|c|c|}\hline
{\bf Parameter} & {\bf Best fit} & {\bf 95$\%$ confidence bound}\\\hline
$a$ & $5.094$ & $(1.691, 8.497)$ \\
$b$ & $1.056$ & $(0.9175, 1.194)$ \\ 
$c$ & $4.476$ & $(-33.29, 42.24)$\\\hline
\end{tabular}

\vspace*{1em}

\noindent 
Estimated parameters of the best power-law fit for the
number of iterations are shown below (adjusted $R^2$ of the estimate
is $0.9983$):

\vspace*{0.5em}

\begin{tabular}{|c|c|c|}\hline
{\bf Parameter} & {\bf Best fit} & {\bf 95$\%$ confidence bound}\\\hline
$a$ &  $0.01109$ & $(0.006155, 0.01602)$ \\
$b$ & $1.356$  & $(1.26, 1.452)$ \\ 
$c$ & $0.6109$ & $(0.4669, 0.755)$\\\hline
\end{tabular}

\vspace*{1em}

\noindent 
The right-hand side of figure~\ref{fig-modeling-percentiles-80} shows
the predictions obtained from the best-fit approximations (shown in
the left) for problems with $n\in(80,200]$. The $95\%$-confidence
prediction bounds are included in the figure (dashed lines). For
example, the $95\%$-confidence upper bound on the population size
for $n=200$ is approximately $1505$, whereas the upper bound on the
number of iterations for $n=200$ is about $16$.

\subsection{Modeling the Distribution Parameters}

The basic idea of this approach to estimating an adequate population
size and the maximum number of iterations is to directly model the
distribution parameters and then predict the distribution parameters
for larger problems. Based on the predicted probability distributions
of the population size and the number of iterations, we can predict
adequate values of these parameters to solve at least a specified
percentage of larger problem instances.

Specifically, we start with the estimated mean and deviation of
the underlying normal distribution for the log-normal fit of the
population size and the number of iterations. The growth of the mean
and the standard deviation is then approximated using an appropriate
function to fit the two estimated statistics.

\begin{figure}
\hspace*{1ex}
{\epsfig{file=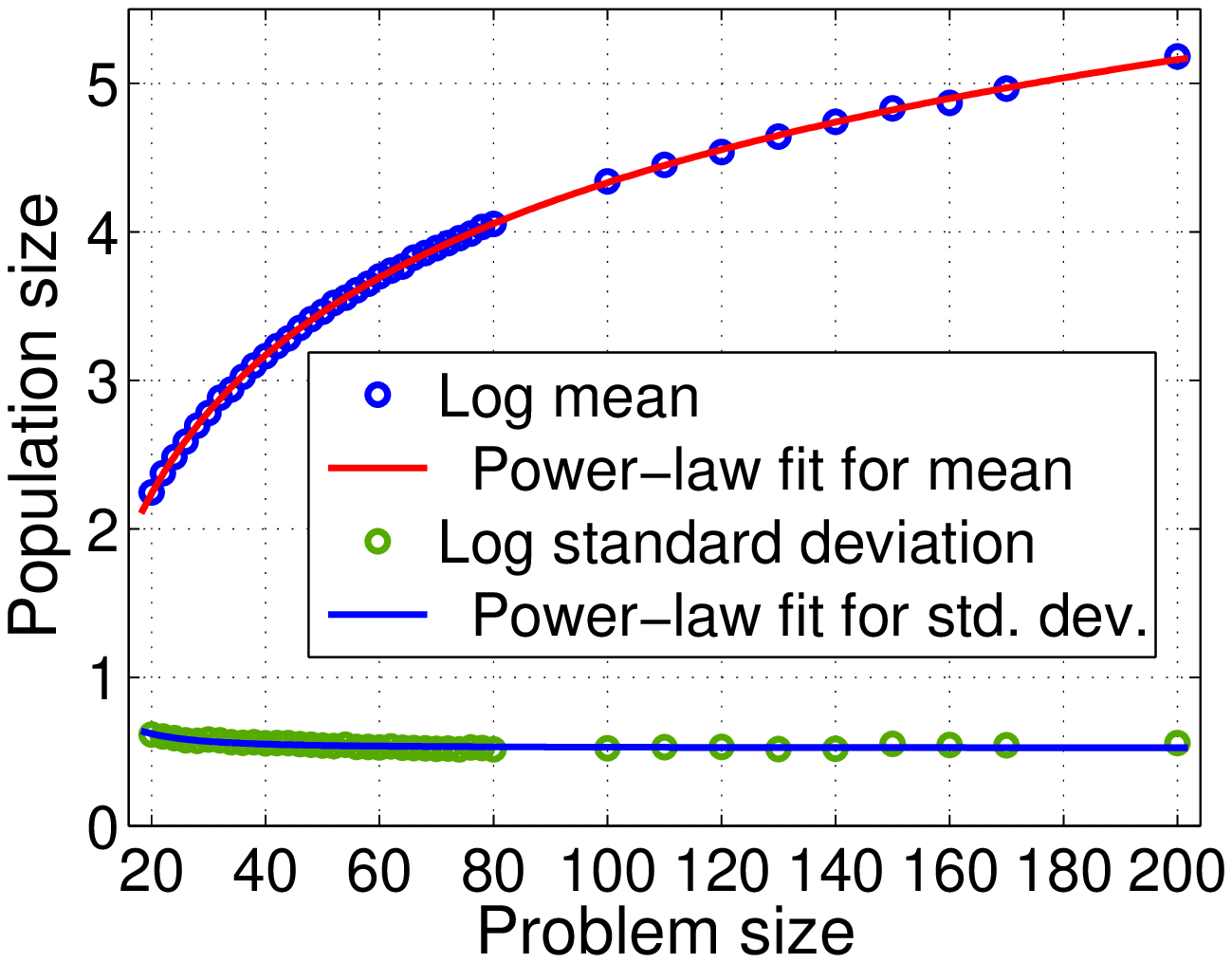,width=0.450\textwidth}}
\hspace*{3ex}
{\epsfig{file=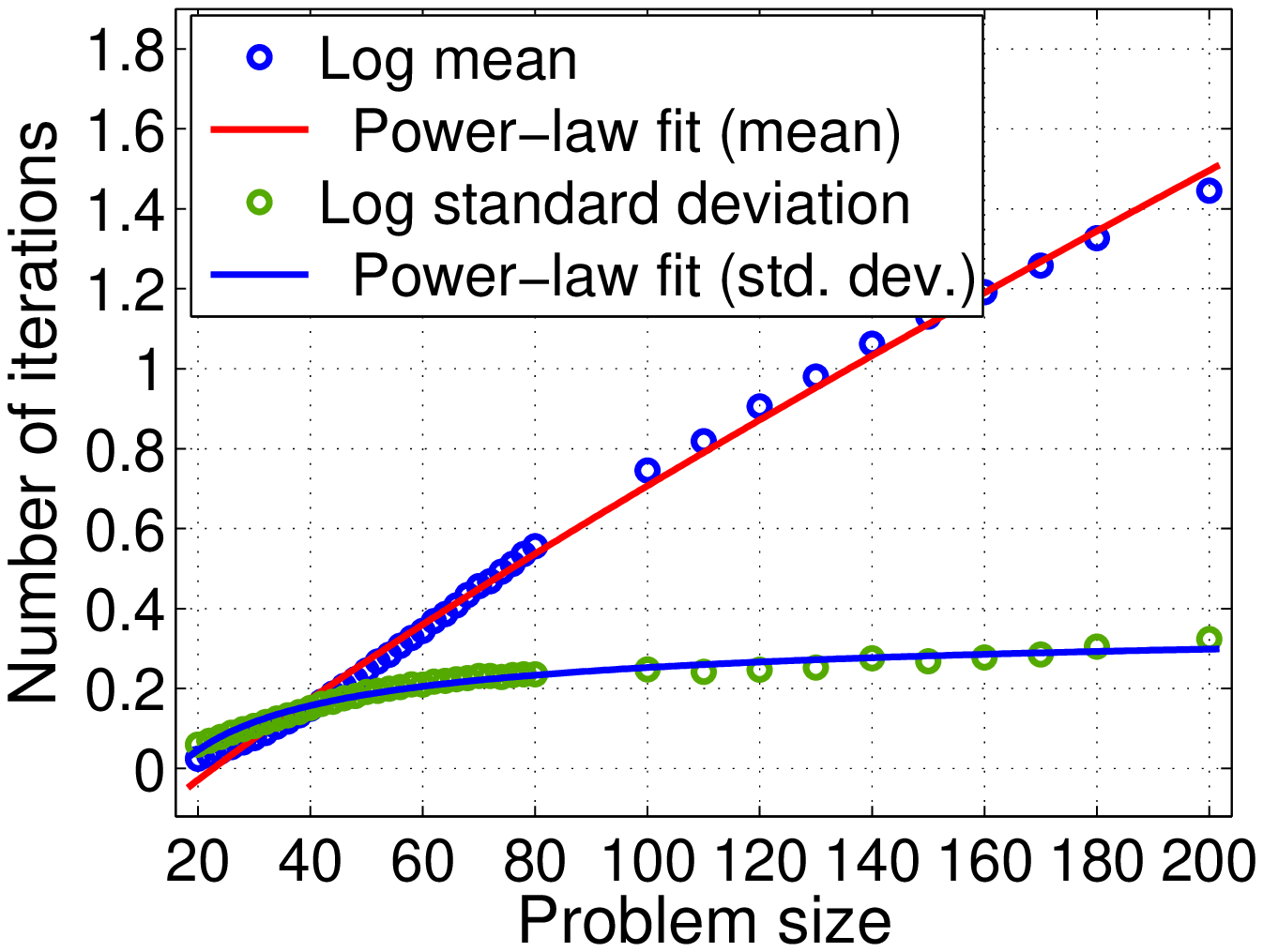,width=0.450\textwidth}}
\caption{
Parameters of the probability distributions governing hBOA parameters
can be modeled directly as shown. The left-hand side displays the
fit for the parameters of the distribution of the population size;
the right hand side shows the fit for the distribution of the number
of iterations. In both cases, the standard deviation appears to be
much more difficult to model than the mean.}
\label{fig-model-dist-params}
\end{figure}

Figure~\ref{fig-model-dist-params} shows the estimated distribution
parameters and the power-law fit for these parameters obtained with
the Matlab curve-fitting toolbox. For both the mean and the standard
deviation, the best match is obtained with the power-law fit. For
the mean, the $R^2$ for the fit is $0.9998$, whereas for the standard
deviation, the $R^2$ is $0.7879$. Thus, for the standard deviation,
the fit is not very accurate. The main reason for this is that the
standard deviation of the underlying normal distribution is rather
noisy and it is difficult to find a model that fits the standard
deviation estimates accurately. Therefore, it appears that the
first approach to predicting the population size and the number of
iterations for larger problems results in more accurate estimates,
although both approaches yield comparable results.

\subsection{Population Doubling}
\label{section-population-doubling}

One of the main features of the problem of finding ground states of
various spin-glass models, including those in two and three dimensions,
is that the problem difficulty varies significantly between different
problem instances. As a result, even the population size and the
number of iterations vary considerably between the different problem
instances. Estimating an upper bound for the population size and
the number of iterations enables us to guarantee that with high
probability, the found spin configurations will indeed be ground
states. Nonetheless, since many problem instances are significantly
simpler than the predicted worst case, this causes us to waste
computational resources on the simple problems. Furthermore, while
in some cases the distribution of the parameters may be relatively
straightforward to estimate accurately, these estimates may be
misleading or difficult to obtain in other cases.

One way to circumvent these problems is to use the following
approach, which is loosely based on the parameter-less genetic
algorithm~\cite{Harik:99a} and the greedy population sizing~\cite{Smorodkina:07}. 
The approach starts with a relatively
small population size $N_{\rm init}$, and executes $num_{\rm runs}$
hBOA runs with that population size (for example, $num_{\rm runs}=10$)
and a sufficient upper bound on the number of iterations. The best
solution of each run is recorded. Next, the procedure is repeated
with the double population size $2N_{\rm init}$, and again the best
solution of each run is recorded. The doubling continues until it
seems unnecessary to further increase the population size, and the
best solution found is then returned.

If a certain population size is too small to reliably identify
the global optimum, we can expect two things to happen: (1)
different runs would result in solutions of different quality (at least
some of these solutions would thus be only locally optimal), and (2)
doubling the population size would provide better solutions. Based on
these observations, we decided to terminate the population-doubling
procedure if and only if all $num_{\rm runs}$ runs end up in the
solution of the same quality and the solution has not improved
for more than $max_{\rm failures}$ rounds of population doubling
(for example, $max_{\rm failures}=2$). Of course, the method can
be tuned by changing the parameters $num_{\rm runs}$ and $max_{\rm
failures}$, depending on whether the primary target is reliability
or efficiency. To improve performance further (at the expense of
reliability), the termination criterion can be further relaxed by
not requiring all runs to find solutions of the same quality.

\noindent
The procedure is summarized in the following pseudo-code (the code
assumes {\em maximization}):
\begin{verbatim}
 find_optimum( N_init, num_runs, max_failures )
 {
   failures=0;
   fold=-infinity;
   fnew=-infinity;
   N=N_init;
   do {
      results = run num_runs runs of hBOA with population size N;
      fnew = best(results);
      if (fold>=fnew)
         failures = failures+1;
      fold = fnew;
      N = N*2;
   } while (best(results)!=worst(results)) or (failures<max_failures);
 }
\end{verbatim}

There are two main advantages of the above procedure for discovering
the global optima. First of all, unlike the approaches based on the
worst-case estimation of the population size, here simpler problems
will indeed be expected to use less computational resources. Second,
we do not have to provide any parameter estimates except for the
maximum number of iterations, which is typically easy to estimate
sufficiently well. Furthermore, even if we do not know how to properly
upper bound the maximum number of iterations, we can use other common
termination criteria. For example, each run can be terminated when
the fitness of the best solution does not improve for a specified
number of iterations or when the fitness of the best solution is
almost equal to the average fitness of the population. Since in many
cases it is difficult to estimate the growth of the population size,
the algorithm presented in this section may be the only feasible
approach out of the three approaches discussed in this paper.

Clearly, there are also disadvantages: Most importantly, if we have
an accurate enough statistical model for the population size and the
number of iterations, modeling the percentiles or parameters of these
distributions allows a highly reliable detection of global optima for
larger problems. On the other hand, if we use the approach based on
doubling the population size, although the termination criteria are
designed to yield reliable results, there are no guarantees that we
indeed locate the global optimum.


\section{Experiments on Larger Problem Instances}
\label{section-experiments-larger}

This section shows the results of applying hBOA to SK instances of
sizes of up to $n=300$. The problems of sizes $n\in[100,200]$ were
solved using parameter estimates created by statistical models of the
percentiles of the estimated distributions of these parameters. Then,
the results on problems of size $n\leq 200$ were used to improve the
model of the population size and the number of iterations, which was
then used to estimate adequate values of the population size and the
number of iterations for problem instances of $n=300$ spins.

\subsection{Solving Instances of $100$ to $200$ Spins}

To solve larger SK spin-glass instances, we first generate $1000$
instances for $n=100$ to $n=200$ spins with step $10$; the number
of instances for each problem size is decreased because as the
problem size grows, it becomes more difficult to reliably locate the
ground states. Then, we use the model of the growth of the $99.999$
percentile of the population size and the number of iterations to
estimate adequate values of these parameters for each value of $n$
with confidence $95\%$. To further improve reliability, for each
problem instance we perform $10$ independent runs and recorded the
best solution found. To verify the final results, for $n=200$, we
make one additional run with hBOA with both parameters twice as big
as those predicted by our model, and compare the obtained results. The
verification does not reveal any inconsistencies and it is thus highly
probable that the obtained spin configurations are indeed true ground
states. Nonetheless, since for $n\geq 100$ we can no longer use branch
and bound, these ground states are not fully guaranteed. Currently,
we are running experiments with the population-doubling approach to
further verify the results obtained for $n\in[100,200]$; as of now,
no inconsistencies have been found, providing further support for the
reliability of the results we have obtained with the previous approach.

After determining the ground states of spin-glass instances for
$n=100$ to $n=200$, we use bisection to find the optimal population
size for hBOA on each instance, similarly as done in the experiments
for $n\leq 80$ (see section~\ref{section-initial-experiments}). Since
for each problem size we generate only 1000 random instances, in
order to obtain more results, we repeat bisection 10 times for each
instance, always with different initial parameters and a different
random seed. Therefore, for each problem instance, we end up with
$100$ successful runs ($10$ successful hBOA runs for each of the
10 bisection runs), and the overall number of successful runs for
each problem size is 100,000. Overall, for problem sizes $n=100$
to $n=200$, we performed 1,100,000 successful runs with hBOA.

\begin{figure}
\hspace*{4.5ex}
{\epsfig{file=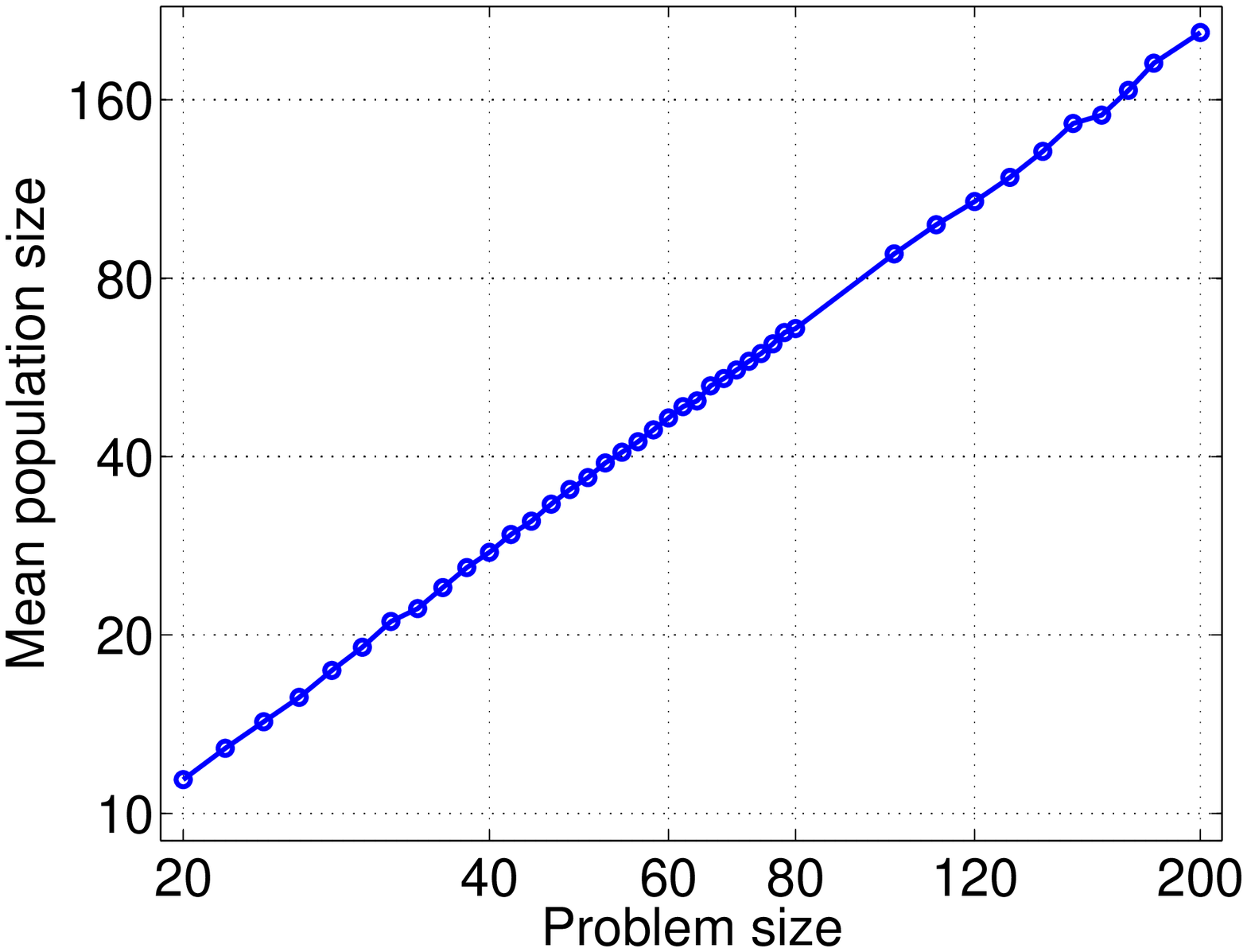,width=0.400\textwidth}}
\hspace*{4.5ex}
{\epsfig{file=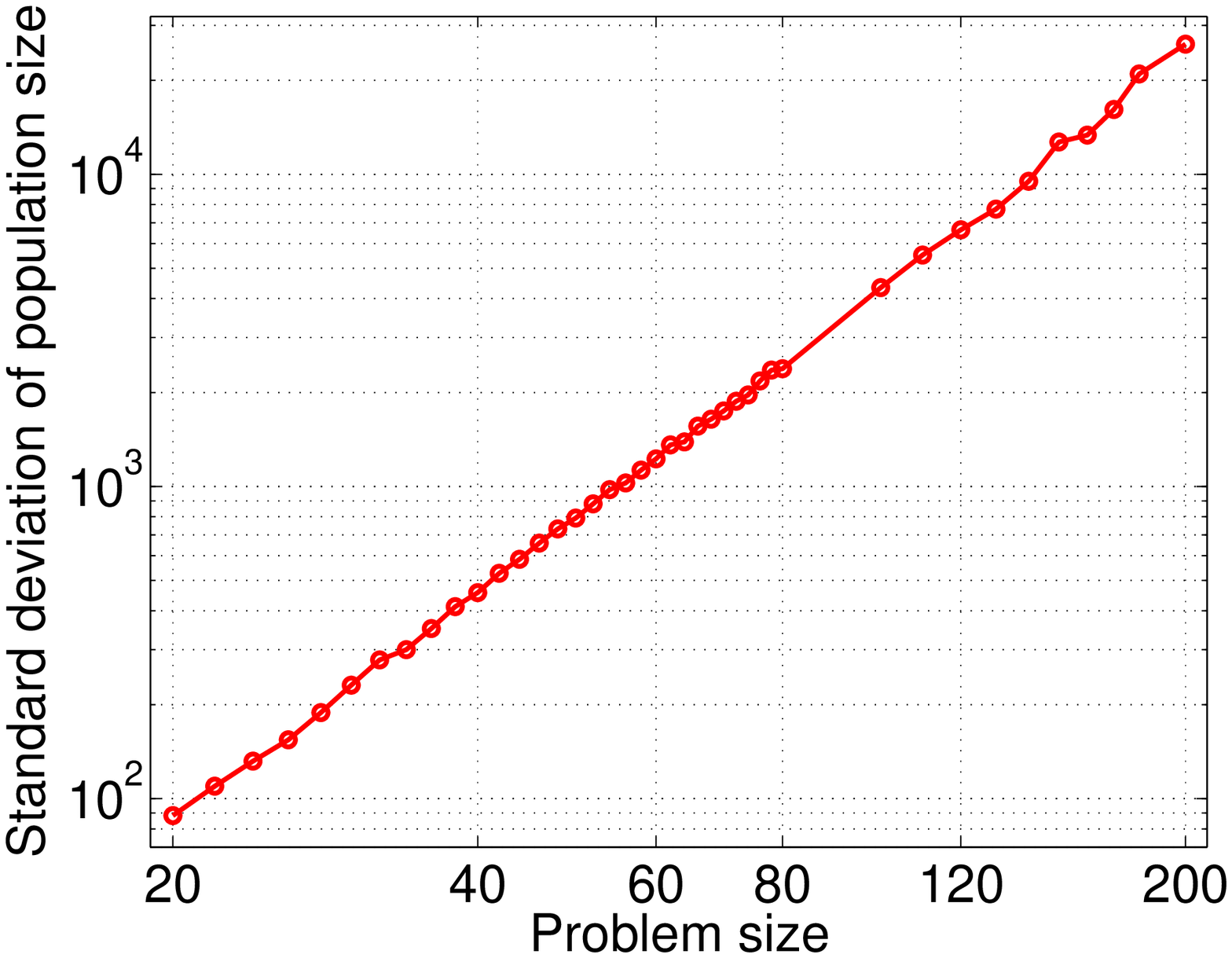,width=0.400\textwidth}}
\hspace*{4.5ex}\\
\hspace*{4.5ex}
{\epsfig{file=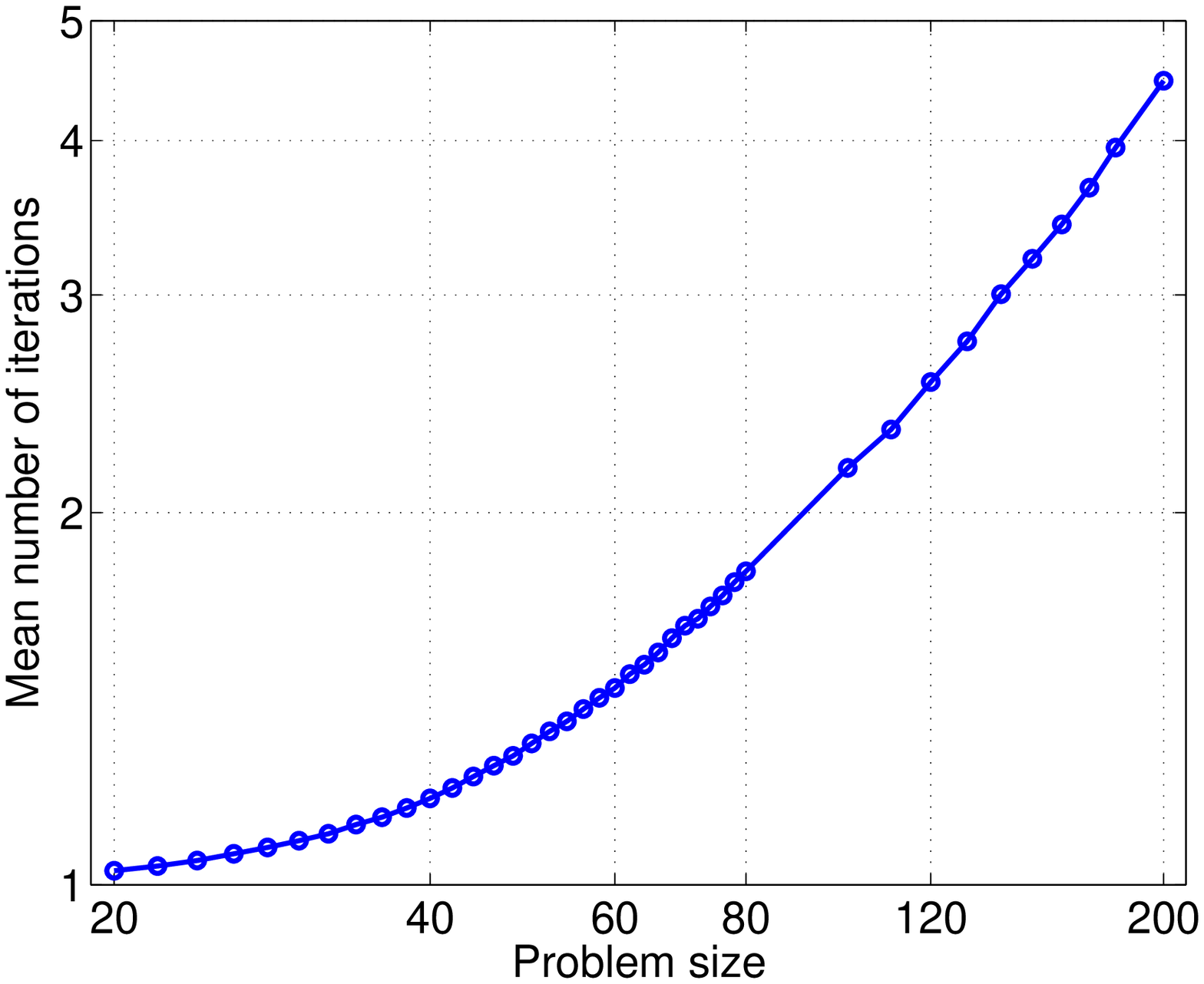,width=0.400\textwidth}}
\hspace*{4.5ex}
{\epsfig{file=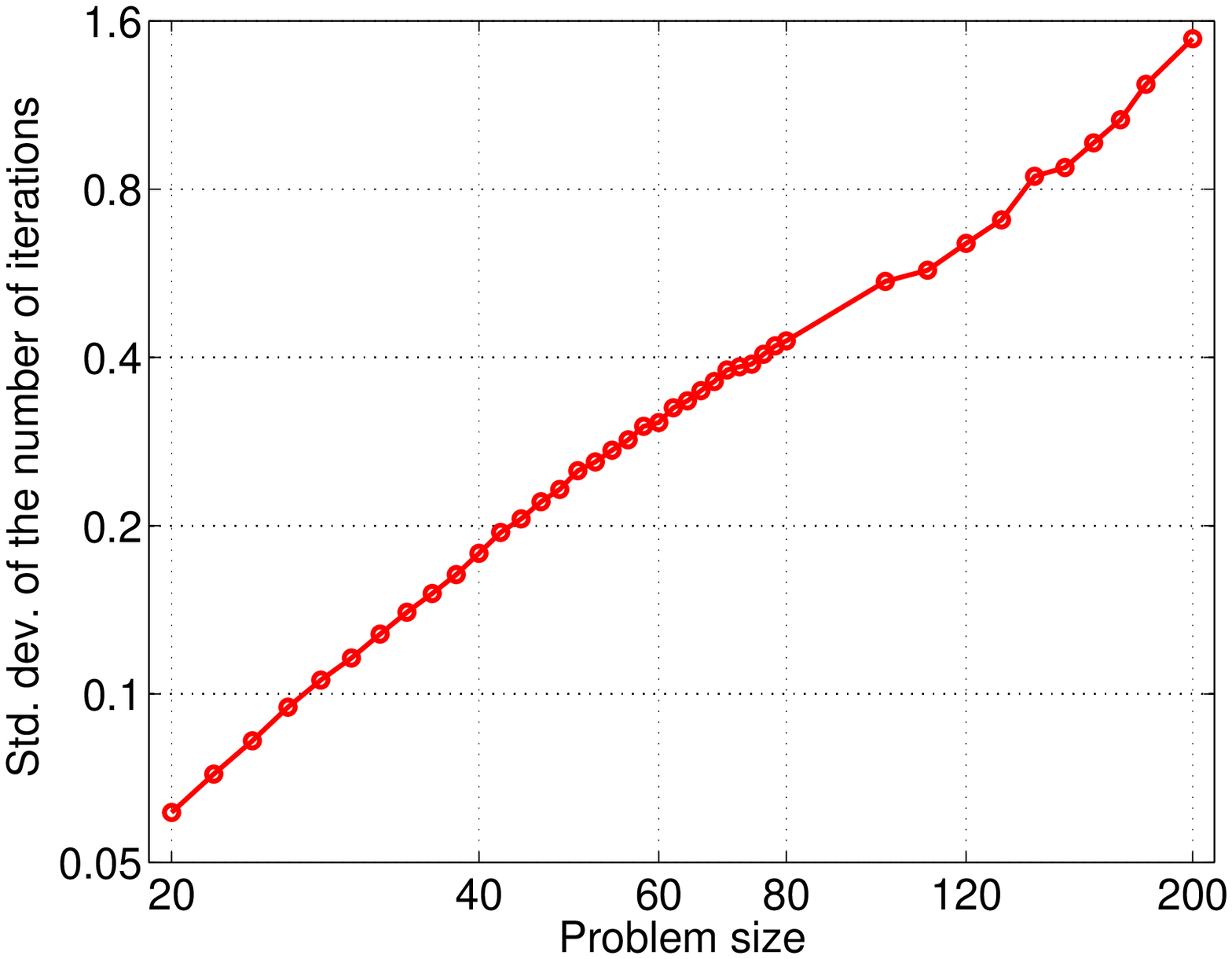,width=0.400\textwidth}}
\hspace*{4.5ex}\\
\hspace*{4.5ex}
{\epsfig{file=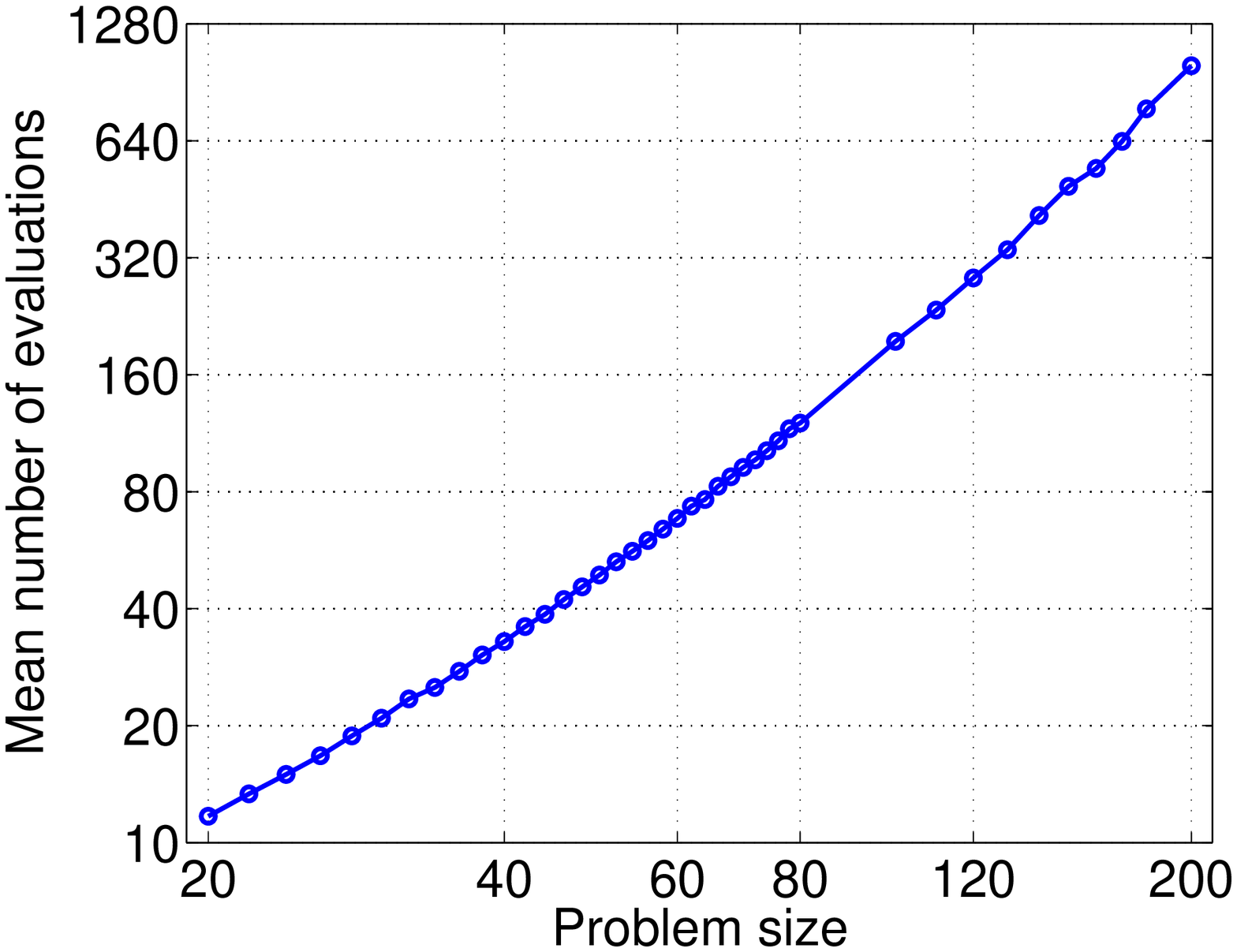,width=0.400\textwidth}}
\hspace*{4.5ex}
{\epsfig{file=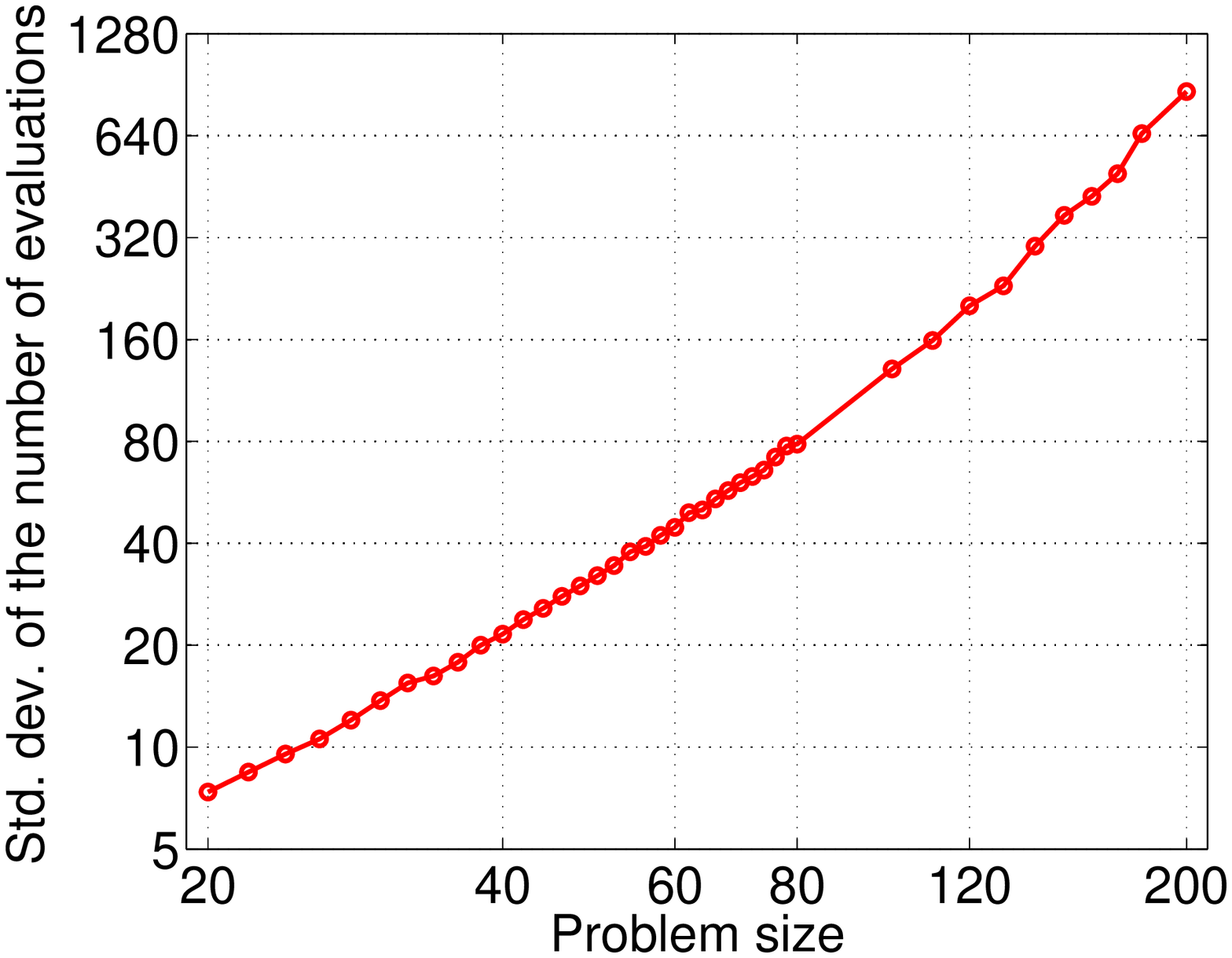,width=0.400\textwidth}}
\hspace*{4.5ex}\\
\hspace*{4.5ex}
{\epsfig{file=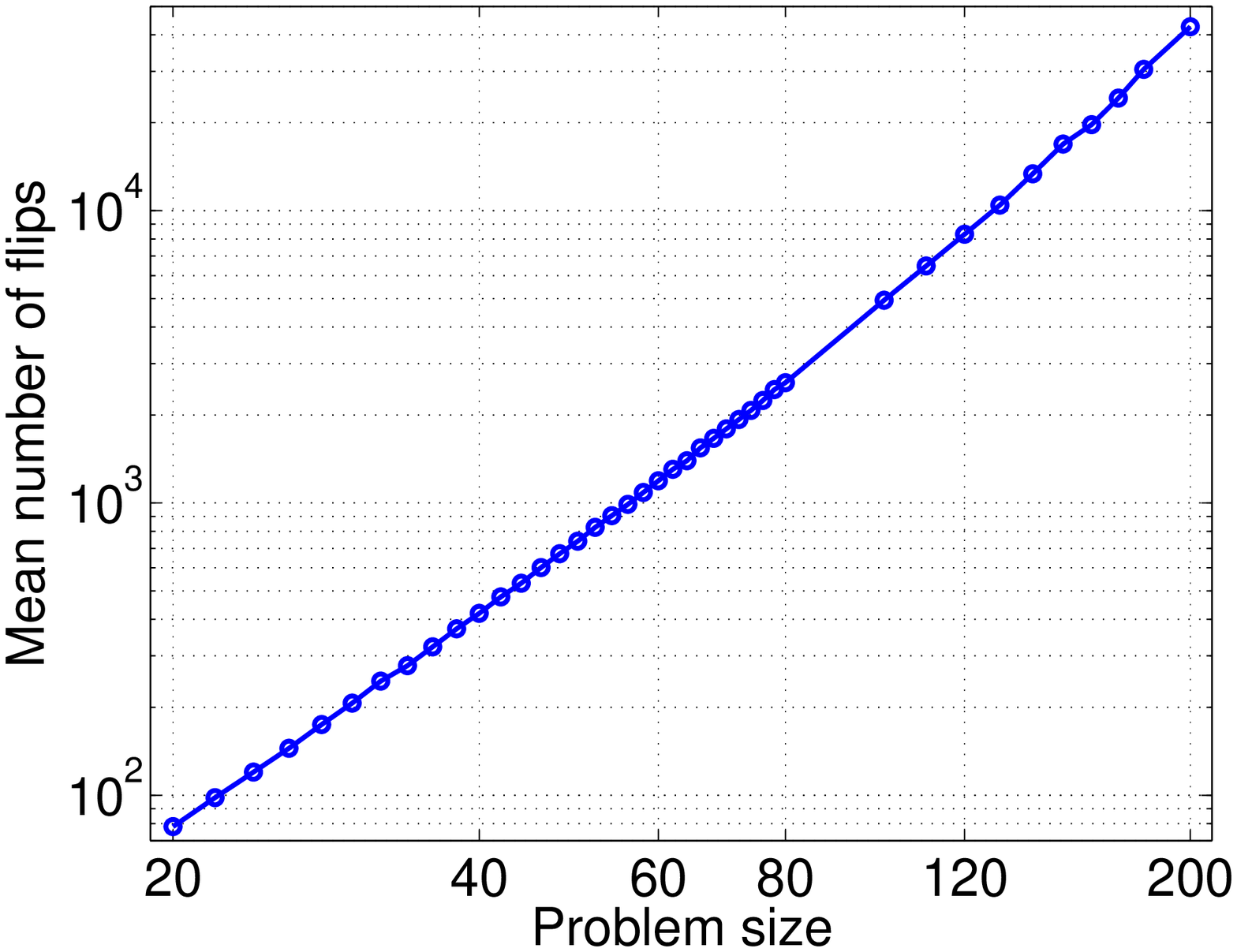,width=0.400\textwidth}}
\hspace*{4.5ex}
{\epsfig{file=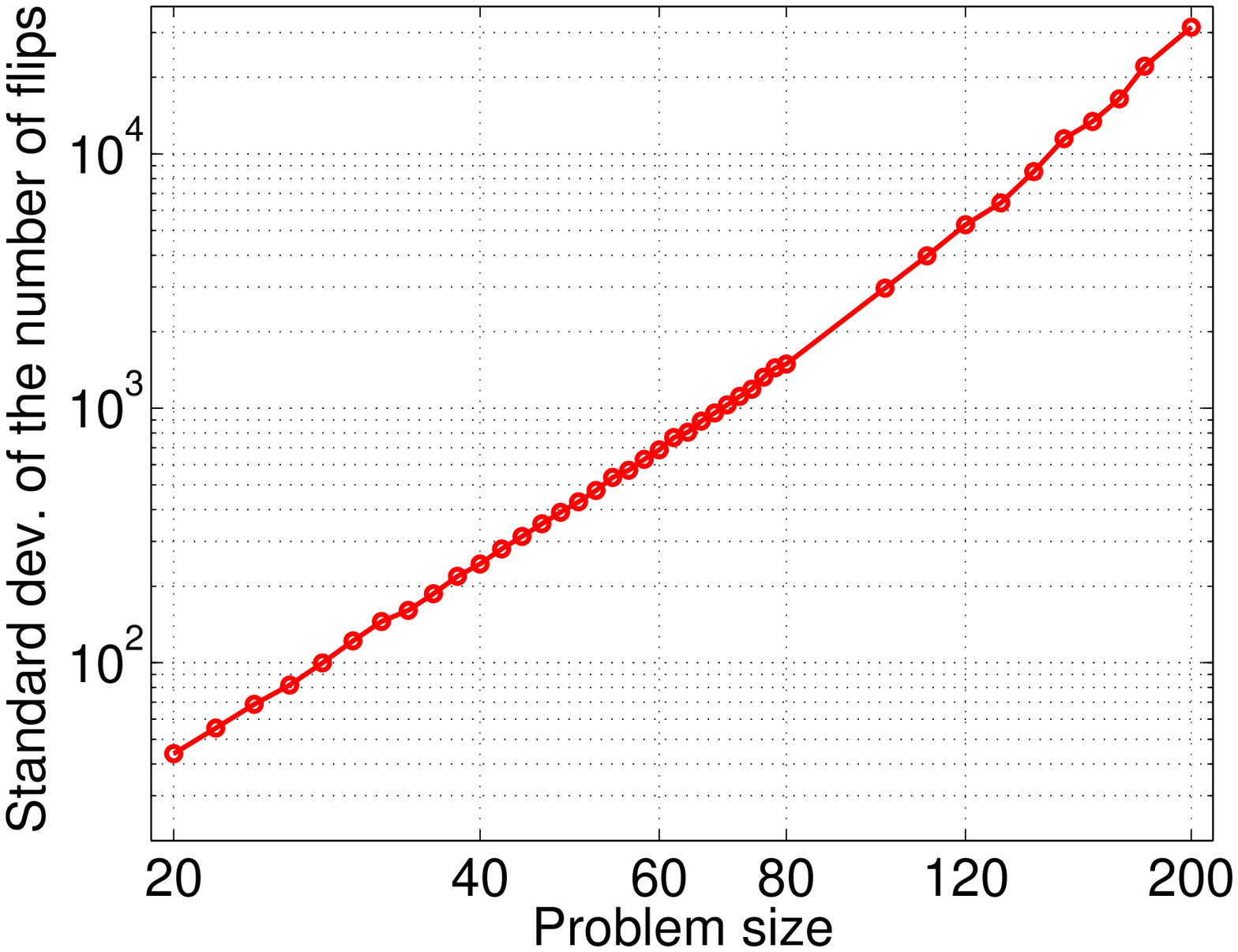,width=0.400\textwidth}}
\hspace*{4.5ex}
\caption{
Mean and standard deviation of the log-normal approximation for the
population size, the number of iterations, the number of evaluations,
and the number of flips for SK spin glasses with $n=20$ to $n=200$.}
\label{fig-mean-stdev-200}
\end{figure}

Analogously to the results on spin-glass instances of sizes $n\leq
80$, we fit the population size, the number of iterations, the number
of evaluations, and the number of flips for each problem size using
log-normal distribution. The resulting mean and standard deviation
for all the statistics is shown in figure~\ref{fig-mean-stdev-200}.

\begin{figure}
\hspace*{1ex}
{\epsfig{file=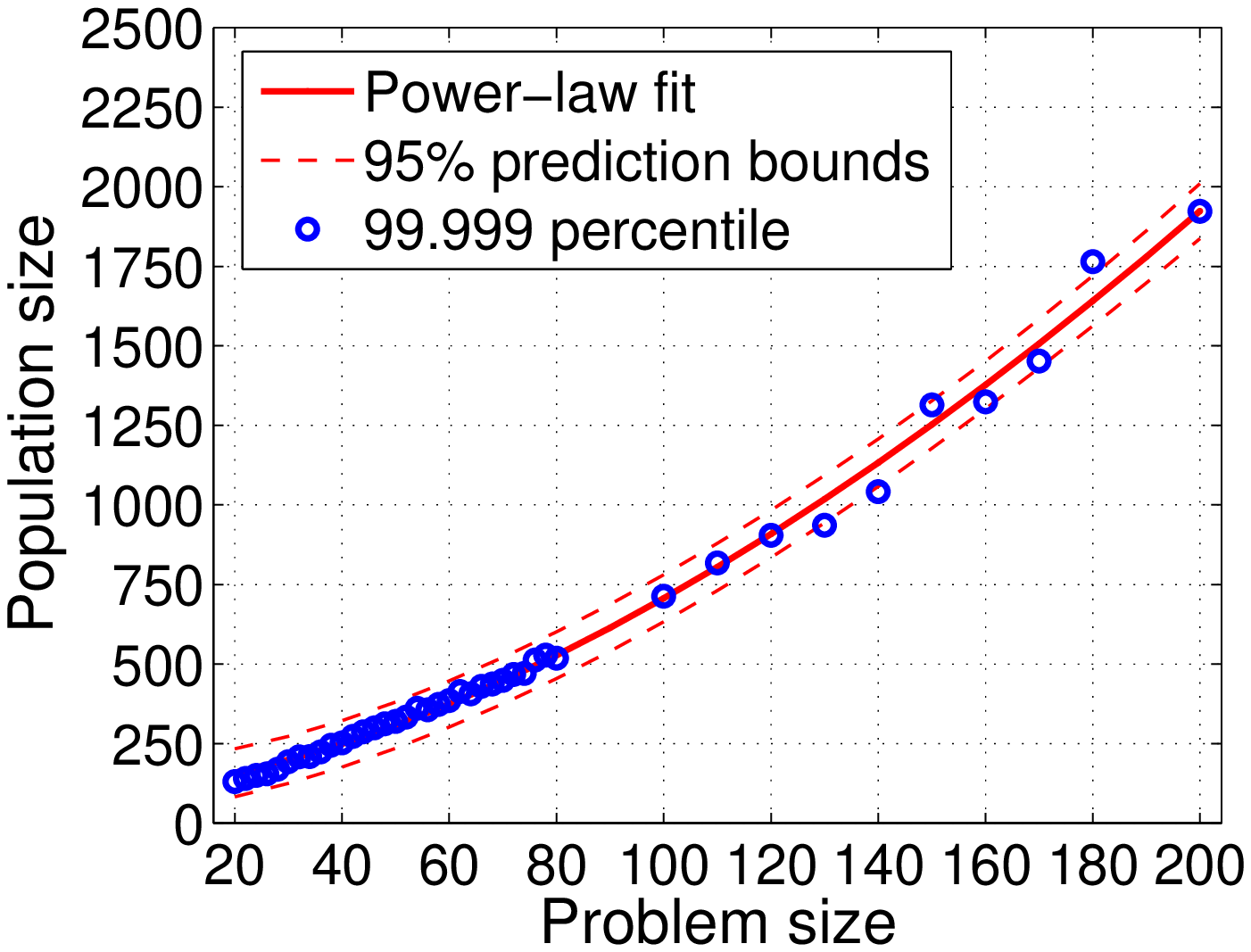,width=0.450\textwidth}}
\hspace*{3ex}
{\epsfig{file=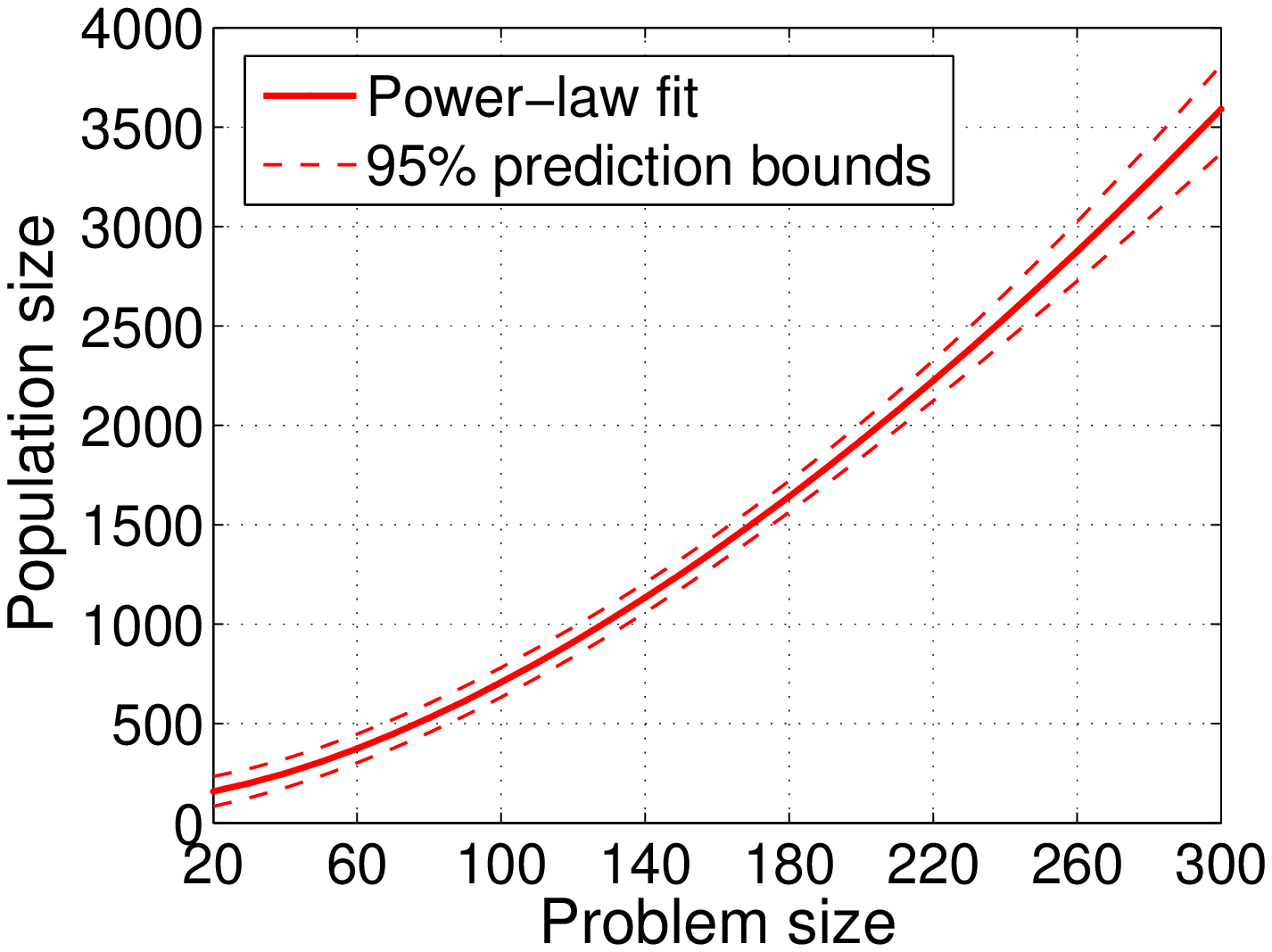,width=0.450\textwidth}}
\\
\hspace*{1ex}
{\epsfig{file=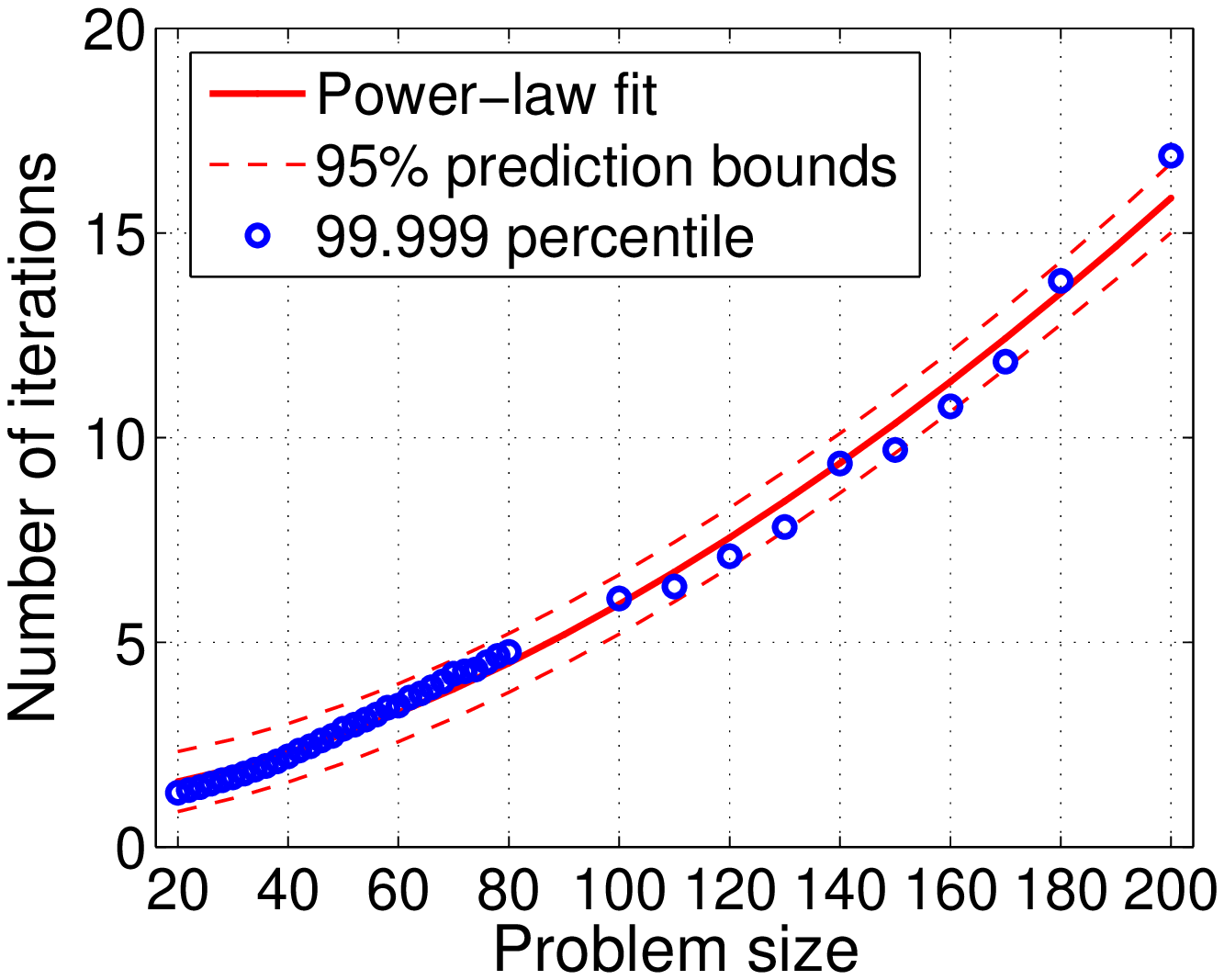,width=0.470\textwidth}}
\hspace*{2.3ex}
{\epsfig{file=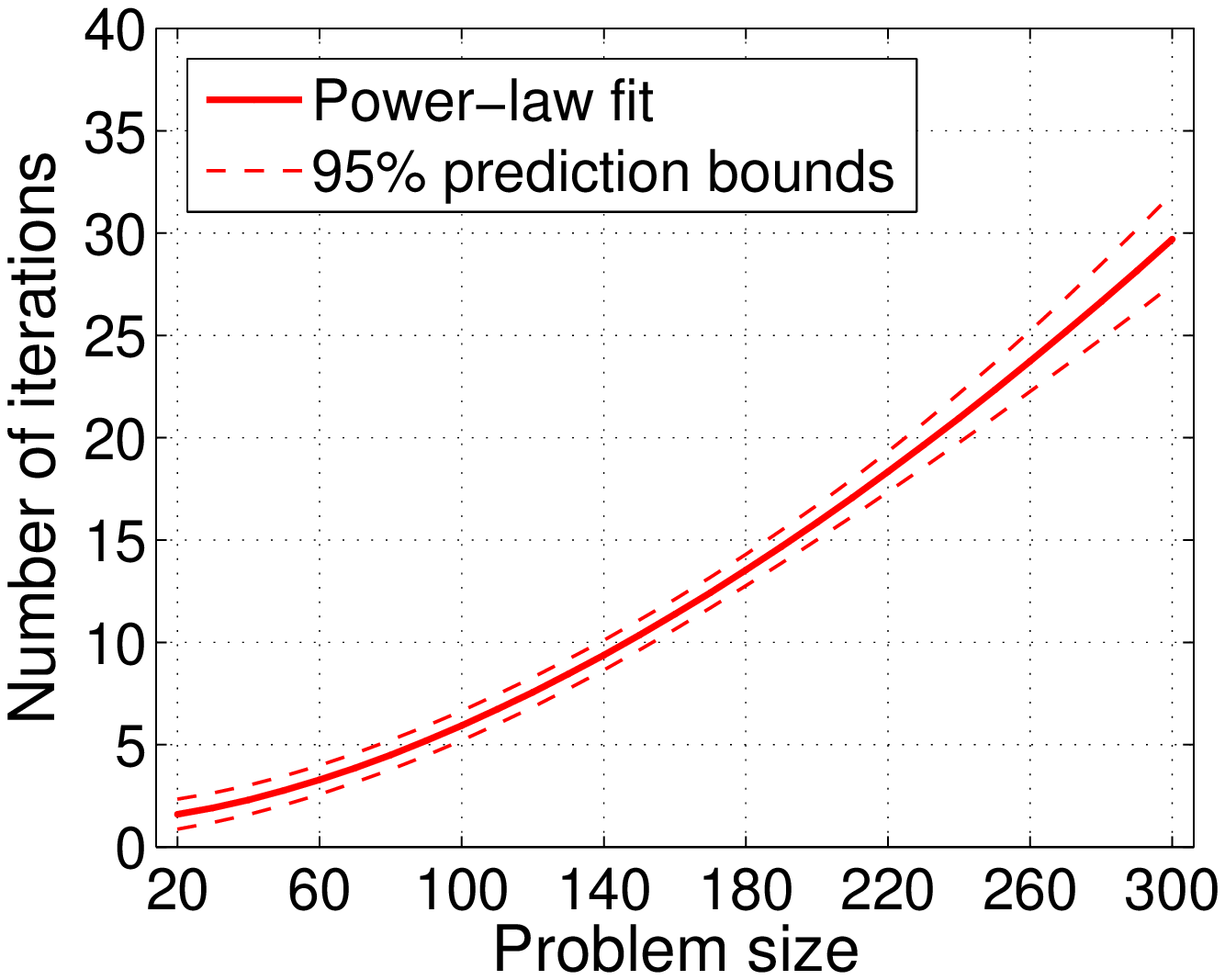,width=0.470\textwidth}}
\caption{
A model of the population size and the number of iterations based on
the growth of the 99.999 percentile of the estimated distributions. The
left-hand side shows the percentiles and the power-law fit which
estimates their growth for $n\leq 200$. The right-hand side shows
the resulting predictions for larger problems of up to $n=300$.}
\label{fig-modeling-percentiles-200}
\end{figure}

An analogous fit to the 99.999 percentile of the population size and
the number of iterations is obtained for problems of sizes $n\leq 200$
with step $2$ from $n=20$ to $n=80$ and with step $10$ from $n=100$ to
$n=200$. The fit is shown in figure~\ref{fig-modeling-percentiles-200}
(left-hand side). The power-law fit performed the best, resulting
in the following parameters for the population size ($R^2$ value for
the fit is $0.9938$):

\vspace*{0.5em}

\begin{tabular}{|c|c|c|}\hline
{\bf Parameter} & {\bf Best fit} & {\bf 95$\%$ confidence bound}\\\hline
$a$ &      $0.3582$ & $(0.1427, 0.5737)$\\
$b$ &        $1.61$ & $(1.496, 1.723)$\\
$c$ &       $113.3$ & $(79.54, 147)$\\\hline
\end{tabular}

\vspace*{1em}

\noindent
The power-law fit for the number of iterations had the following
parameters ($R^2$ value for the fit is $0.9909$):

\vspace*{0.5em}

\begin{tabular}{|c|c|c|}\hline
{\bf Parameter} & {\bf Best fit} & {\bf 95$\%$ confidence bound}\\\hline
$a$ &    $0.002379$ & $(0.0006242, 0.004133)$\\
$b$ &       $1.646$ & $(1.506, 1.786)$\\
$c$ &       $1.264$ & $(0.945, 1.584)$\\\hline
\end{tabular}

\vspace*{1em}

\noindent 
We use the above model to predict adequate values of the population
size and the number of iterations for $n=300$, $3812.44$ for
the population size and $30.1702$ for the number of iterations,
respectively. The predictions for problems $n\leq 300$ are shown in
figure~\ref{fig-modeling-percentiles-200} (right-hand side).

\subsection{Solving Instances of $300$ Spins}

To solve even larger problem instances, we generate $1000$ random
instances of the SK spin glass of $n=300$ bits. These instances are
then solved using hBOA with the population size set according to the
upper limit estimated in the previous section from problems of sizes
$n\leq 200$. Similarly as in the experiments for $n\in[100,200]$, hBOA
is run 10 times on each problem instance and the best solution found
in these 10 runs is recorded. Then, we run bisection 10 times on each
problem instance, resulting in 100 successful runs for each problem
instance (10 successful hBOA runs for each of the 10 bisection runs).

After analyzing the distribution of the various statistics collected
from hBOA runs on spin-glass instances of $n=300$ spins, it became
clear that while for smaller problems, the log-normal distribution
provided an accurate and stable model of the true distribution, for
$n=300$, the fit with the log-normal distribution does no longer seem
to be the best option and the distribution is more accurately reflected
by the generalized extremal value distribution. This is surprising,
since the log-normal fits for smaller problems are very accurate
and they provide more stable results than the generalized extreme
value fits. As presented in reference~\cite{billoire:07} the thermodynamic
limiting behavior of the SK model is only probed for $n \gtrsim 150$
spins. Interestingly, this threshold agrees qualitatively with the
change of the fitting functions.

To verify our previous estimates based on the log-normal distribution,
we repeated the same procedure with the generalized extreme value
distribution. This resulted in somewhat larger upper bounds on the
population size. We are currently verifying all problem instances
with these new estimates for $n\in[100,300]$. Furthermore,
we are running the approach based on population doubling (see
section~\ref{section-population-doubling}) for all instances with
$n\in[100,300]$ so that we have further support for the global
optimality of the spin configurations found with the initial
approach. So far, we haven't found any instances for which a
better solution would be found with larger populations or the
population-doubling approach.


\section{Comparison of hBOA and GA}
\label{section-experiments-comparison}

This section compares the performance of hBOA, GA with two-point
crossover and bit-flip mutation, and GA with uniform crossover and
bit-flip mutation. All algorithms are configured in the same way
except for the population size, which is obtained separately for
each problem instance and each algorithm using the bisection method
described earlier in the paper.

To compare algorithms `$A$' and `$B$,' we first compute the ratio of
the number of evaluations required by $A$ and $B$, separately for each
problem instance. Then, we average these ratios over all instances of
the same problem size. If the ratio is greater than $1$, the algorithm
$B$ requires fewer evaluations than the algorithm $A$; therefore, with
respect to the number of fitness evaluations, we can conclude that
$B$ is better than $A$. Similarly, if the ratio is smaller than $1$,
then we can conclude that with respect to the number of evaluations,
$A$ performs better than $B$. Analogous ratios are also computed
for the number of flips required until the optimum has been reached,
which provides us with an even more important measure of computational
complexity than the number of evaluations.

\begin{figure}
\hspace*{1ex}
{\epsfig{file=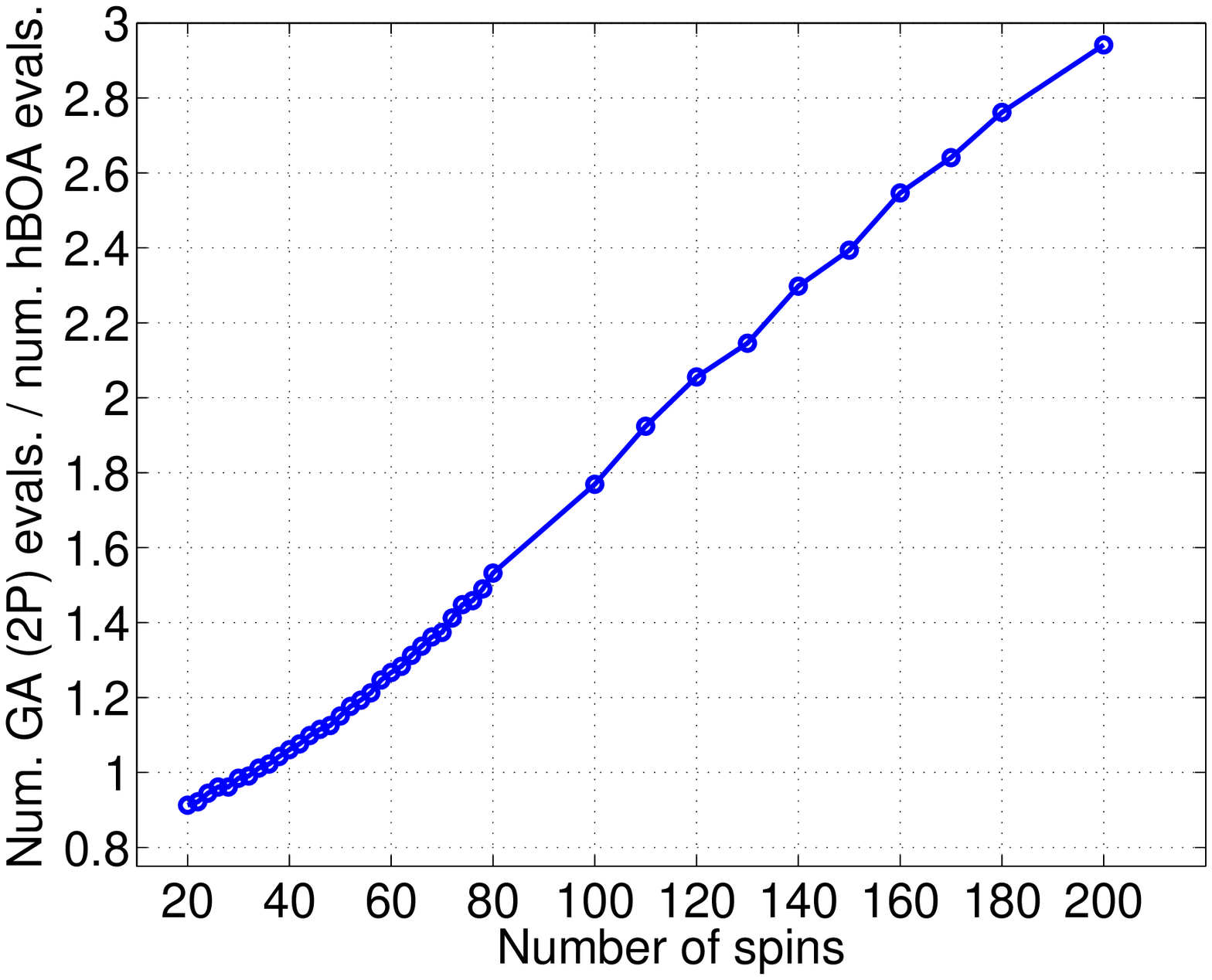,width=0.450\textwidth}}
\hspace*{3ex}
{\epsfig{file=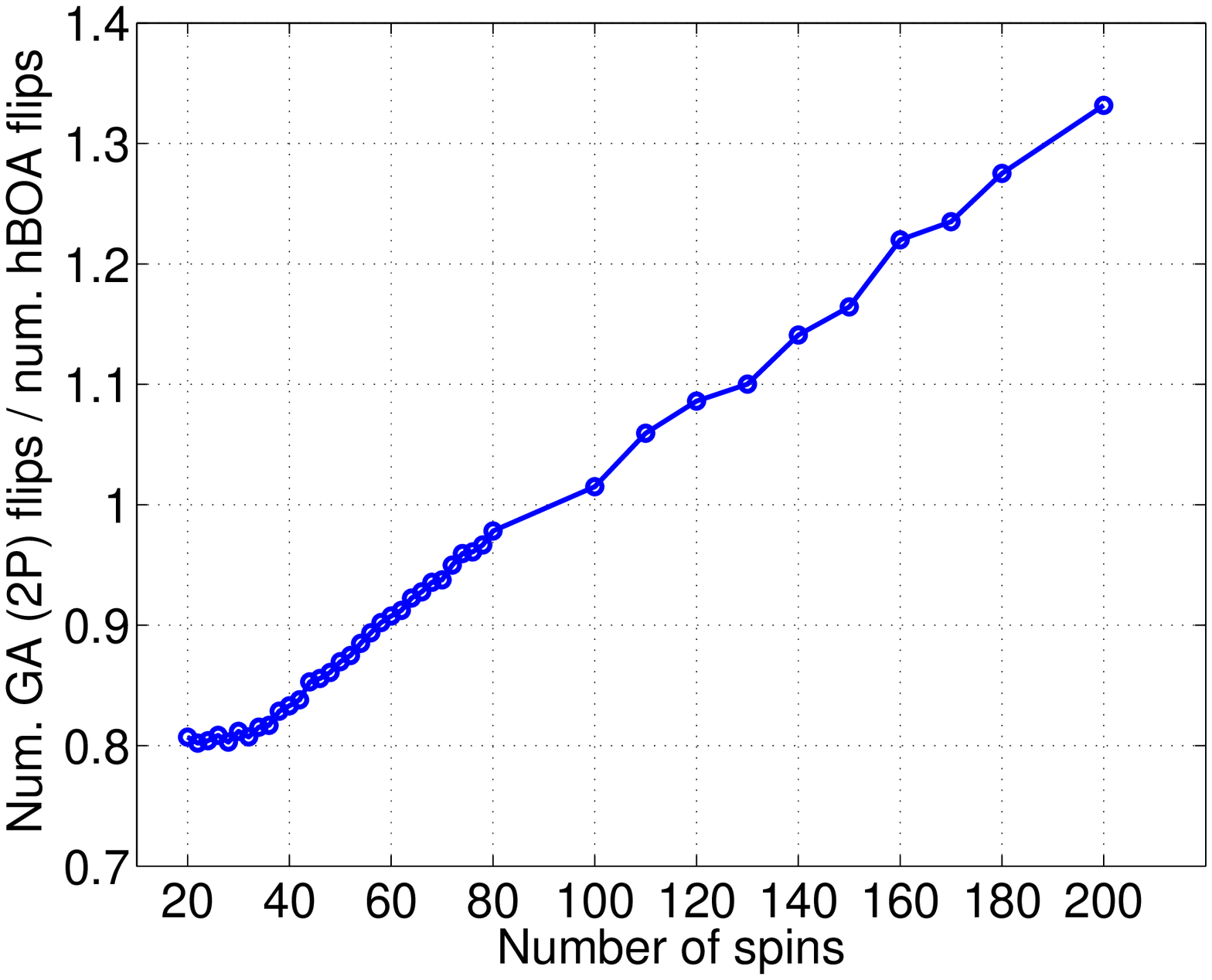,width=0.450\textwidth}}
\\
\hspace*{1ex}
{\epsfig{file=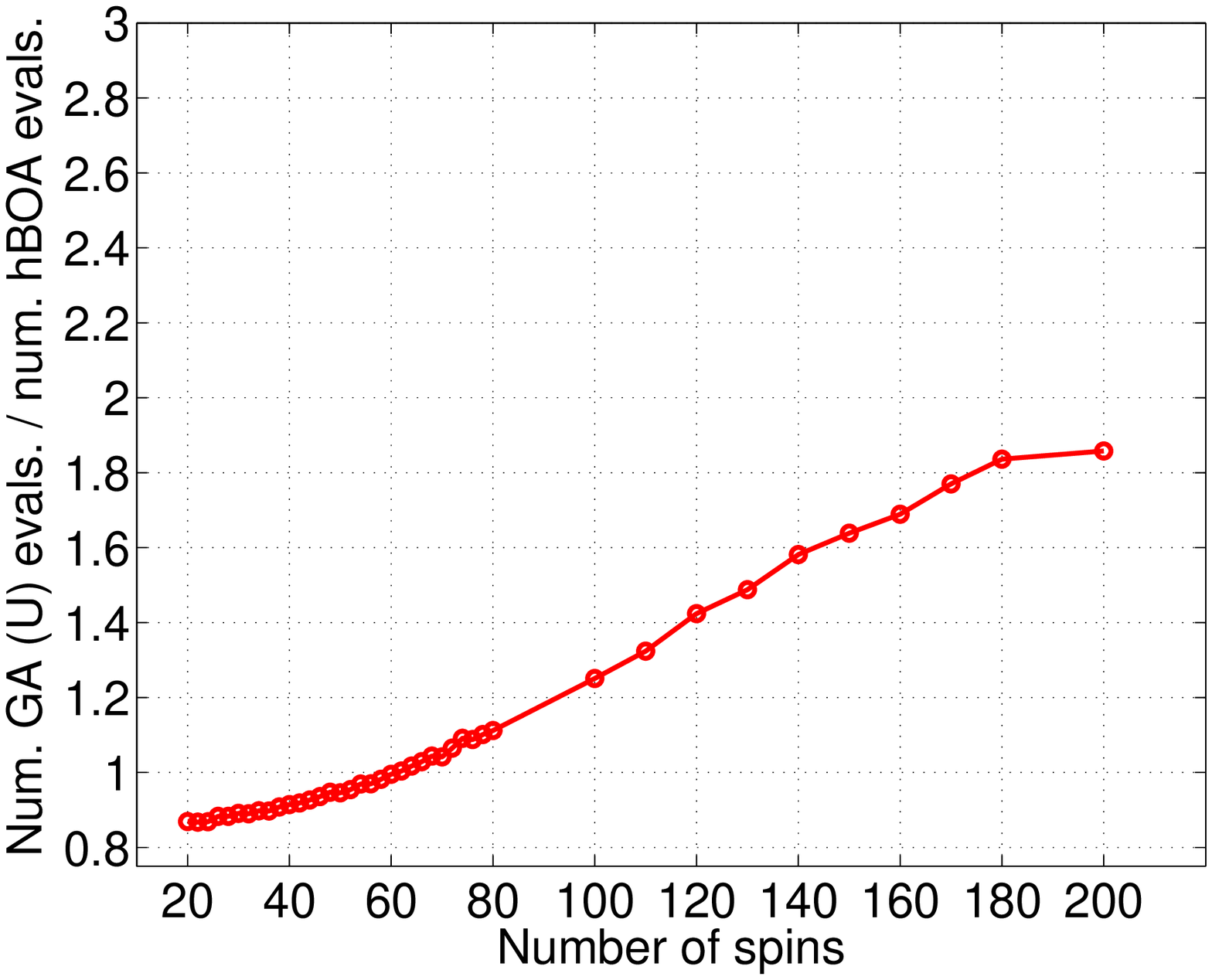,width=0.450\textwidth}}
\hspace*{3ex}
{\epsfig{file=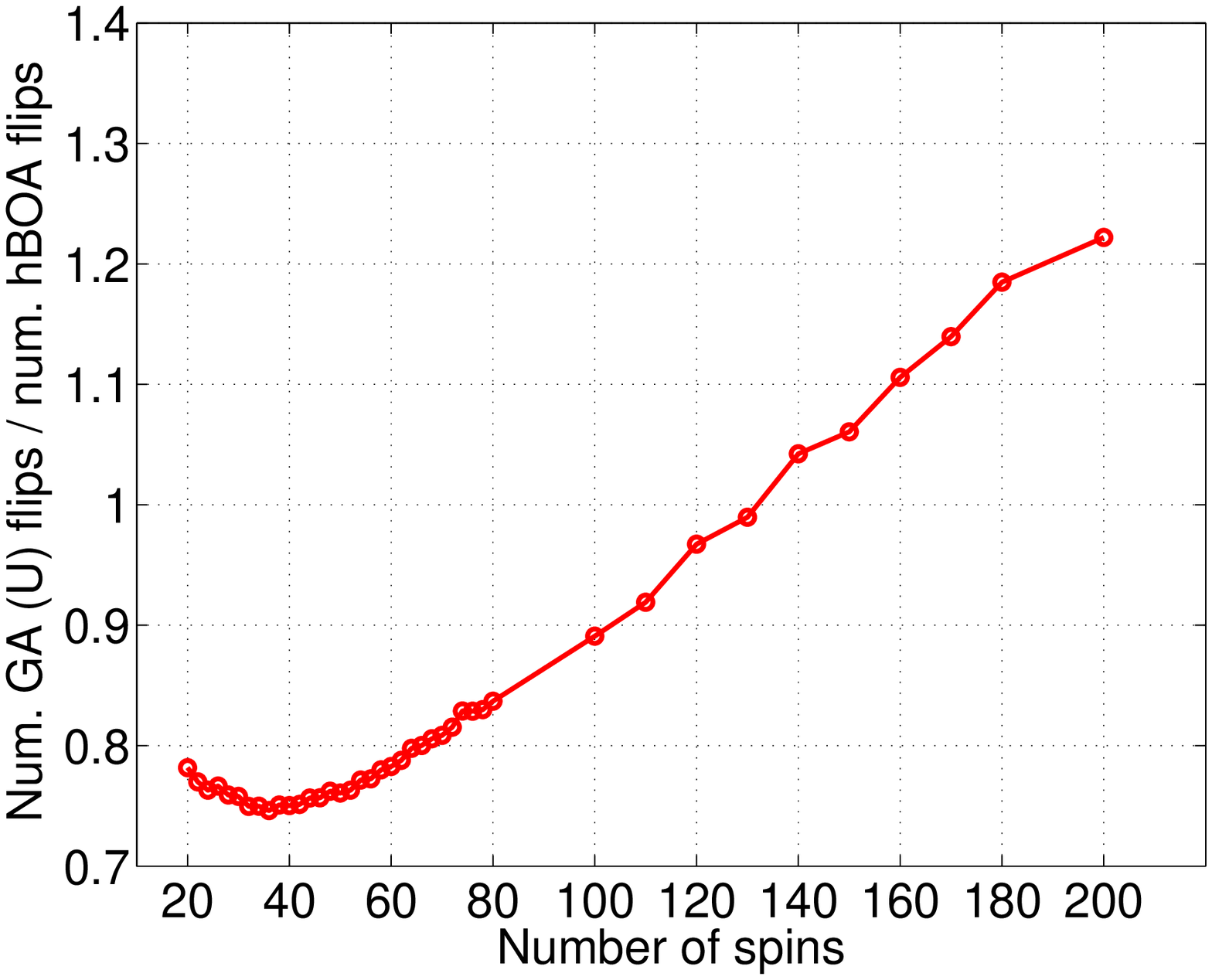,width=0.450\textwidth}}
\\
\hspace*{1ex}
{\epsfig{file=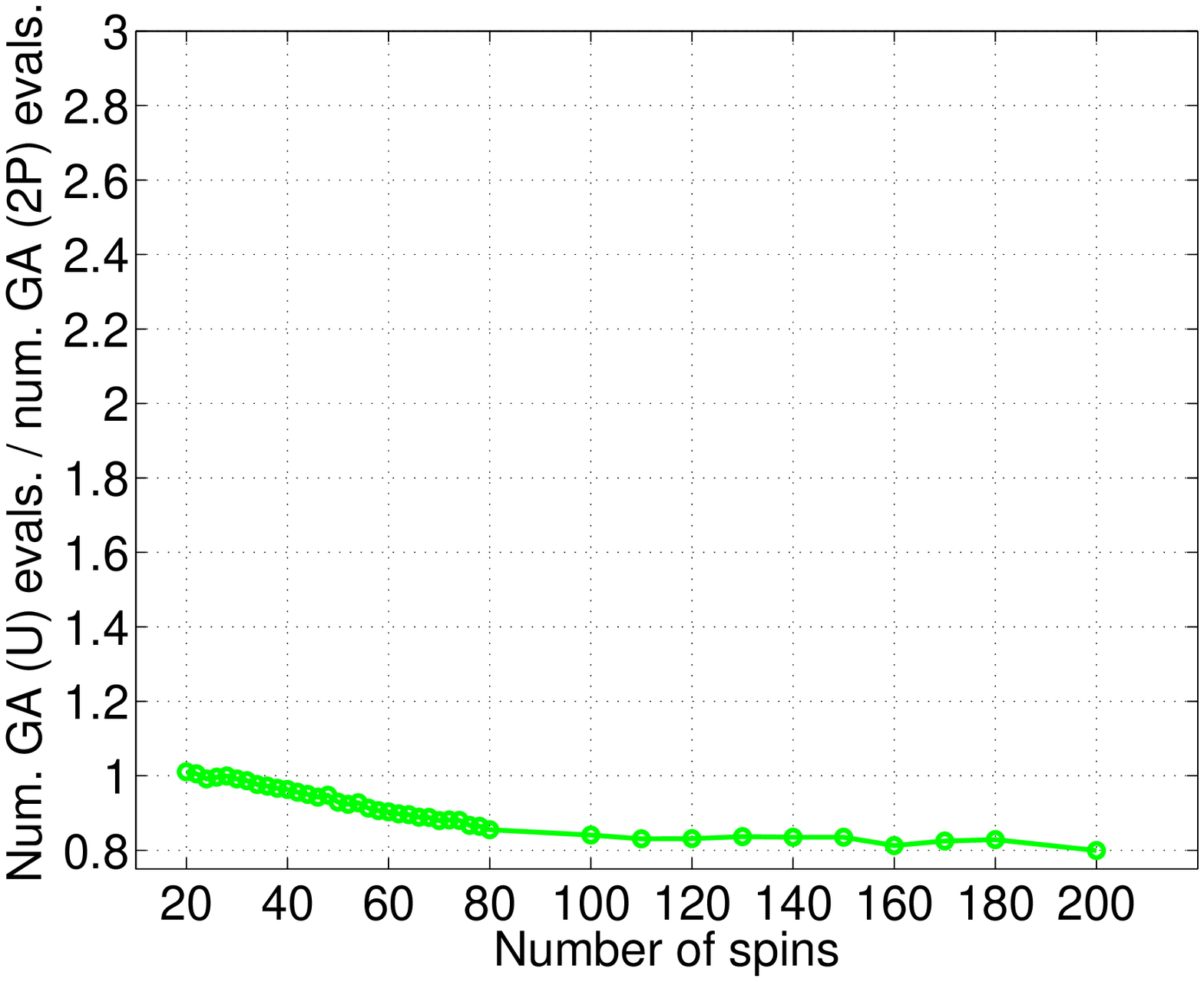,width=0.450\textwidth}}
\hspace*{3ex}
{\epsfig{file=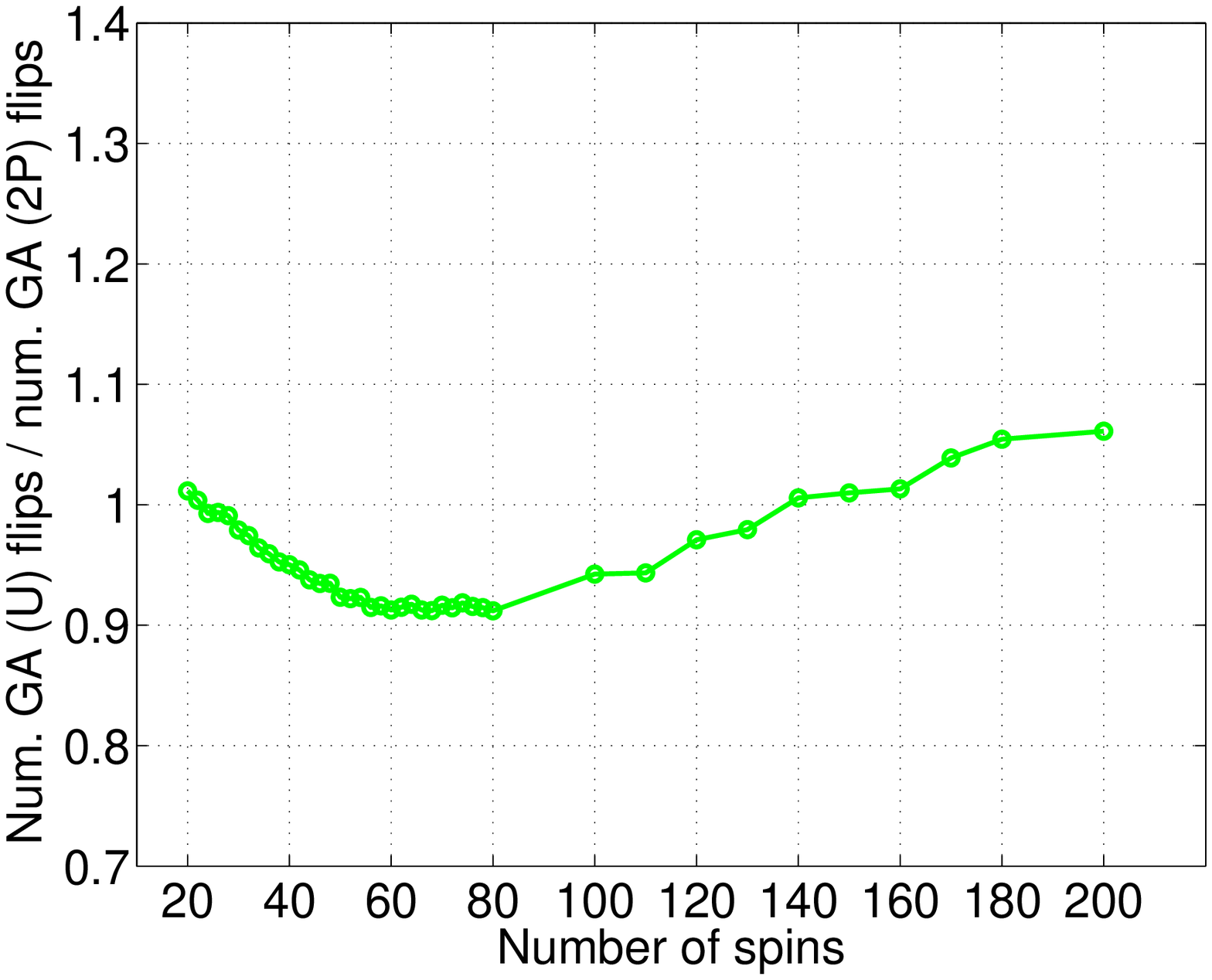,width=0.450\textwidth}}
\\
\caption{
Comparison of hBOA, GA with uniform crossover, and GA with two-point
crossover with respect to the overall number of evaluations and
the number of flips of the local searcher. The relative performance
is visualized as the ratio of the number of evaluations (number of
flips) required by pairs of compared algorithms. The results clearly
show that the factor by which hBOA outperforms GA (with respect to
both the number of evaluations and the number of flips) grows with
problem size. Since the number of flips is a more important measure of
overall complexity than the number of evaluations, uniform crossover
is expected to outperform two-point crossover as the problem size
increases.}
\label{fig-comparison}
\end{figure}

The results of pair-wise comparisons between all three algorithms are
shown in figure~\ref{fig-comparison}. The results clearly indicate that
hBOA outperforms both GA variants and the gap between hBOA and the GA
variants increases with problem size. Therefore, for larger problems,
the performance differences can be expected to grow further. From the
comparison of the two GA variants, it is clear that while with respect
to the number of evaluations, two-point crossover performs better,
with respect to the number of flips, uniform crossover performs better
with increasing problem size.

While the differences between hBOA and GA are not as significant
for problem sizes considered in this work, since the gap between
these algorithms grows with problem size, for much larger problem,
the differences can be expected to become significant enough to
make GA variants intractable on problems solvable with hBOA in
practical time. 

\section{Future Work}
\label{section-future-work}

The most immediate milestone to tackle is to further increase the
size of the systems for which we can reliably identify ground states
with the ultimate goal of obtaining global optima for SK instances
significantly larger than $10^3$ spins. One approach is to directly
use the methods discussed in this paper and incrementally increase
problem size. Although hBOA performs very well on the systems we
have tested so far, since the problem of finding ground states
of SK spin-glass instances is NP-complete, the time complexity
is expected to grow very fast, especially on the most difficult
problem instances. That is why one of the important precursors of
future success in achieving this goal is to incorporate existing
efficiency enhancement techniques~\cite{Sastry:06,Pelikan:06b} into
hBOA and design new efficiency enhancement techniques tailored to
the SK spin-glass model. One of the most promising directions for
improving hBOA performance is hybridization, where instead of the
simple deterministic hill climber, more advanced techniques can be
incorporated; this can significantly increase the size of problems
solvable by hBOA with practical time complexity, done similarly for
2D and 3D spin glasses~\cite{Pelikan:06}.

Another interesting topic for future work is to use problem instances
obtained in this research to test other optimization algorithms and
compare their performance with that of hBOA and GA. Good
candidates for such a comparison are exchange Monte Carlo (parallel
tempering) \cite{hukushima:96} adapted for the ground-state 
search \cite{Moreno:03,katzgraber:04c,koerner:06}, genetic algorithm
with triadic crossover~\cite{Pal:96}, extremal 
optimization~\cite{Boettcher:05}, and hysteric optimization~\cite{Pal:06}. 
While some of these algorithms were argued to solve relatively large SK spin-glass 
instances, there is no guarantee that the results obtained represent true 
ground states and the performance of these methods on certain classes of 
the SK spin-glass model is relatively poor. Nonetheless, a rigorous 
comparison of these algorithms with the techniques studied in this paper 
remains an important topic for future work. 

Finally, this work can be extended to other interesting types of
instances of the SK spin-glass model and other difficult combinatorial
problems. One of the extensions is to consider other distributions
of couplings, where in particular, bimodal distributions are of
interest due to the high degeneracy of the ground state. Testing
hBOA to see if the method delivers all possible configurations for a
given ground-state energy with the same probability is of paramount
importance \cite{Moreno:03}.  Another interesting extension of
the model is to impose a distance metric between pairs of spins and
modify the coupling distribution based on the distance between the two
connected spins~\cite{Katzgraber:07}. The latter has the advantage
that, while the connectivity of the model is kept constant, the range
of the interactions can be tuned continuously between a system in
a mean-field universality class to a short-range nearest-neighbor
model.  This provides an ideal benchmark for optimization algorithms
in general.


\section{Summary and Conclusions}
\label{section-conclusions}

This paper applied the hierarchical Bayesian optimization algorithm
(hBOA) and the genetic algorithm (GA) to the problem of finding ground
states of instances of the Sherrington-Kirkpatrick (SK) spin-glass
model with Ising spins and Gaussian couplings, and analyzed performance
of these algorithms on a large set of instances of the SK model. First,
10,000 random problem instances were generated for each problem size
from $n=20$ to $n=80$ with step $2$ and ground states of all generated
instances were determined using the branch-and-bound algorithm. Then,
hBOA was applied to these instances, and its parameters and performance
were analyzed in detail. Since problems of $n\geq 100$ spins are
intractable with branch and bound, we proposed several approaches
to reliably identifying ground states of such problem instances
with hBOA. One of the proposed approaches was applied to problem
instances of sizes $n\in[100,300]$ (1000 random instances for each
problem size). Analogous experiments as with hBOA were also performed
with the genetic algorithm (GA) with bit-flip mutation and two common
crossover operators. Performance of hBOA and the two GA variants was
compared, indicating that hBOA outperforms both GA variants and the
gap between these two algorithms increases with problem size.


Our study presents for the first time a detailed study of genetic and
evolutionary algorithms applied to the problem of finding ground states of the SK
spin-glass model. The lessons learned and the techniques developed
in tackling this challenge should be important for optimization
researchers as well as practitioners.


\section*{Acknowledgments}

The authors would like to thank Kumara Sastry for helpful discussions
and insightful comments.

This project was sponsored by the National Science Foundation
under CAREER grant ECS-0547013, by the Air Force Office of
Scientific Research, Air Force Materiel Command, USAF, under grant
FA9550-06-1-0096, and by the University of Missouri in St. Louis
through the High Performance Computing Collaboratory sponsored by
Information Technology Services, and the Research Award and Research
Board programs.

The U.S.  Government is authorized to reproduce and distribute
reprints for government purposes notwithstanding any copyright notation
thereon. Any opinions, findings, and conclusions or recommendations
expressed in this material are those of the authors and do not
necessarily reflect the views of the National Science Foundation, the
Air Force Office of Scientific Research, or the U.S. Government. Some
experiments were done using the hBOA software developed by Martin
Pelikan and David E. Goldberg at the University of Illinois at
Urbana-Champaign and most experiments were performed on the Beowulf
cluster maintained by ITS at the University of Missouri in St. Louis.

H.G.K.~would like to thank the Swiss National Science Foundation for
financial support under grant No.~PP002-114713. Part of the simulations
were performed on the Gonzales cluster of ETH Z\"{u}rich.

\bibliographystyle{abbrv}

\end{document}